\documentclass{article}
\usepackage{iclr2027_conference,times}
\usepackage{graphicx}
\usepackage{float}
\usepackage{placeins}
\usepackage{needspace}
\usepackage{amsmath,amssymb}
\usepackage{multirow}
\usepackage{booktabs}
\usepackage{array,tabularx}
\usepackage{longtable}
\usepackage{listings}
\usepackage[hidelinks]{hyperref}
\usepackage{url}
\hypersetup{
  pdftitle={HistCAD: Constraint-Aware Parametric CAD Histories for Evaluating Editability},
  pdfauthor={Xintong Dong, Chuanyang Li, Peng Zheng, Chuqi Han, Jiaxin Jing, Hailong Shen, Yanzhi Song, Zhouwang Yang},
  pdfsubject={Constraint-aware parametric CAD histories and editability evaluation},
  pdfkeywords={HistCAD, parametric CAD, sketch constraints, editability}
}

\providecommand{\Description}[1]{}

\newcommand{\widefigurewidth}{\linewidth}

\title{
 HistCAD: Constraint-Aware Parametric CAD Histories for Evaluating Editability
}

\author{%
\parbox[t]{\dimexpr\textwidth-2\tabcolsep\relax}{\raggedright%
\textbf{\mbox{Xintong Dong}, \mbox{Chuanyang Li}, \mbox{Peng Zheng}, \mbox{Chuqi Han}, \mbox{Jiaxin Jing}, \mbox{Hailong Shen}, \mbox{Yanzhi Song\thanks{Co-corresponding authors.}}\hspace{0.4em} \& \mbox{Zhouwang Yang\footnotemark[1]}}%
}\\
University of Science and Technology of China, Hefei, Anhui, P. R. China\\
{\normalfont \href{mailto:weiss@mail.ustc.edu.cn}{weiss@mail.ustc.edu.cn}, \href{mailto:lichuanyang@mail.ustc.edu.cn}{lichuanyang@mail.ustc.edu.cn}, \href{mailto:pengzheng@mail.ustc.edu.cn}{pengzheng@mail.ustc.edu.cn}}\\
{\normalfont \href{mailto:hanchuqi@mail.ustc.edu.cn}{hanchuqi@mail.ustc.edu.cn}, \href{mailto:smallmatch@mail.ustc.edu.cn}{smallmatch@mail.ustc.edu.cn}, \href{mailto:shenhail@mail.ustc.edu.cn}{shenhail@mail.ustc.edu.cn}}\\
{\normalfont \href{mailto:yanzhis@ustc.edu.cn}{yanzhis@ustc.edu.cn}, \href{mailto:yangzw@ustc.edu.cn}{yangzw@ustc.edu.cn}}
}

\iclrfinalcopy 

\begin{document}
\maketitle
\lhead{Preprint}
\rhead{}

\begin{abstract}
Sketch constraints specify geometric conditions for constructing and modifying parametric
CAD models. We study whether predicted constraints allow a given history to reproduce the
required initial model and support prescribed dimensional edits. We introduce
\mbox{HistCAD}, an executable representation and dataset whose Academic and Industrial
collections contain 180{,}495 parametric construction histories with entity-referenced sketch
constraints and retained feature operations. Predictors receive these histories with the geometry and feature definitions retained and
explicit sketch constraints removed. They
generate constraints for every sketch without seeing the edit request. The benchmark
compares models built with alternative constraint sets for the same history under the same
dimensional edit. An edit succeeds when the model reproduces the required initial geometry,
reaches the target value, preserves specified relations and unedited dimensions in the target
sketch, and rebuilds through the complete history. A predictor trained on both collections
and supplied with descriptions of the input histories achieves overall edit success of
52.4\% on Academic and 29.0\% on Industrial. Models retaining only endpoint-connectivity
constraints in the target sketch and the history's constraints elsewhere can reach the target
and rebuild while failing preservation. For all-sketch predictions, we retain the target-sketch
prediction and restore the history's constraints in other sketches. More models then reproduce
the required initial geometry, and some of these newly matched models complete the edit.
HistCAD connects constraint learning to the construction and revision of parametric CAD models.
\end{abstract}

\begin{figure}[!htbp]
\centering
\includegraphics[width=\widefigurewidth]{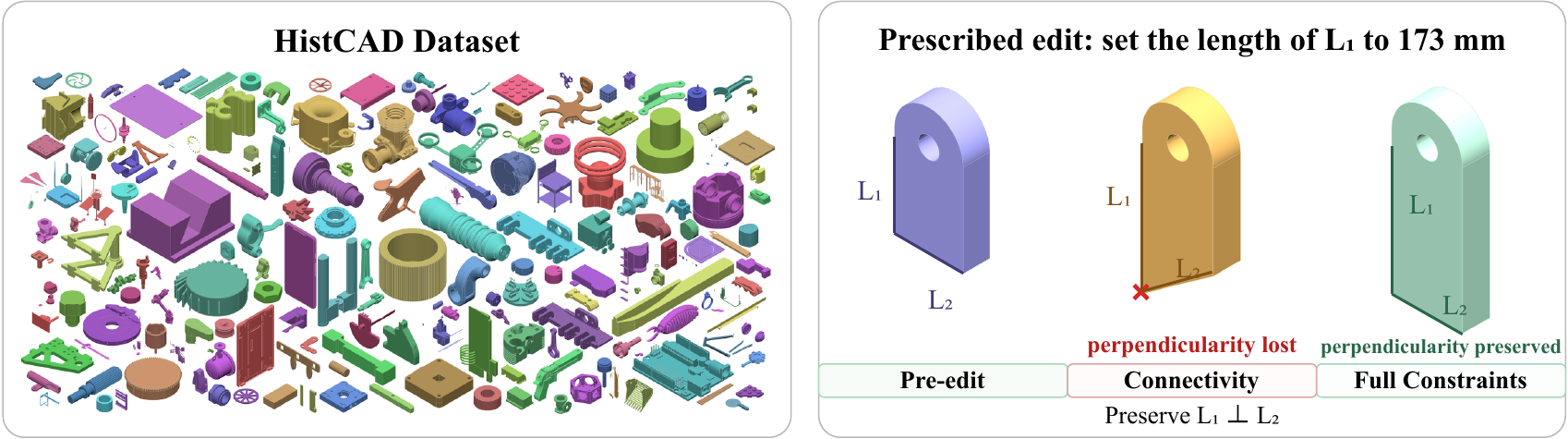}
\caption{HistCAD geometry examples (left) and an illustrative dimensional edit (right). In the illustrated sketch, Full retains all constraints, whereas Connectivity retains only profile-connectivity constraints. Both CAD models start from the same geometry, reach the target length, and rebuild successfully, but only the Full variant preserves the specified perpendicularity \(L_1\perp L_2\).}
\Description{HistCAD geometry examples appear beside an illustrative dimensional edit. The edit changes L1's length and requires L1 to remain perpendicular to L2. Two CAD models with matching pre-edit geometry and different sketch constraints receive the same request. Both attain the target length and rebuild successfully, but only the model under Full Constraints preserves perpendicularity.}
\label{fig:central-counterexample}
\end{figure}

\section{Introduction}
\label{sec:intro}

Parametric CAD supports design revision by updating related geometry when dimension values
change~\citep{camba2016reusability}. Within a sketch, geometric constraints specify relations
between entities, while dimensional constraints set sizes, angles, and
distances~\citep{ault1999using,company2020constraints}. For example, an arc can be constrained
to share its center with a circle and remain tangent to a sketch line. If the perpendicular
distance from that center to the line increases, the arc's radius must also increase to
maintain both relations. The constraints require these relations to hold in the edited geometry. Thus, models with identical initial geometry but different sketch constraints
can yield different geometries under the same dimensional edit.

We study whether predicted sketch constraints allow a history to reproduce the required
initial model and support a requested dimensional change while preserving specified relations
and unedited dimensions. The sketch solver applies the predicted constraints to compute the geometry
used by feature operations to build the initial and edited CAD models. Studying these
outcomes therefore requires data that identify the entities involved in each constraint and
retain the feature operations that use the solved geometry.

We introduce HistCAD, an executable representation and dataset of 180{,}495 parametric
construction histories with geometric and dimensional sketch constraints. The Academic
collection combines the extrusion-based histories from DeepCAD~\citep{wu_2021_iccv} and
Fusion~360~\citep{willis2020fusion}, while Industrial adds native histories with operations
beyond extrusion. Each constraint identifies the sketch entities or points it acts on.
These entity-referenced constraints provide supervision for generating a constraint set for
every sketch. The input history retains the sketch geometry and feature operations and omits explicit
sketch constraints. The predictor is not given the edit request. Predictions can replace the history's constraint sets while the initial primitive definitions,
feature parameters, geometric references, and operation order remain fixed.

We develop an executable benchmark that compares models built with alternative constraint
sets for the same history under the same dimensional edit. For each edit, we specify the
target value and the geometric relations and unedited dimensions to preserve in the target
sketch. The model built with the history's own constraints provides the required initial
geometry. Predicted constraints are used in both initial construction and edited replay.
Different constraint sets can produce different solved geometry from the same initial
primitives, so each model must reproduce the required initial geometry before the edit is
applied. After editing, the solved target sketch must reach the requested value and preserve
all specified relations and unedited dimensions. The complete construction history must also
rebuild successfully. The models in Figure~\ref{fig:central-counterexample} both reach the
target and rebuild, but only one preserves the specified perpendicularity.

All-sketch constraint prediction achieves up to 52.4\% overall edit success on Academic and
29.0\% on Industrial. Target-sketch controls show that reaching the requested dimension and
rebuilding validly can occur without preserving all specified relations and unedited
dimensions. Restoring the history's constraints in other sketches, with the target-sketch
prediction fixed, allows more models to reproduce the required initial geometry.

\noindent\textbf{Contributions.}
We introduce an executable representation and dataset that retain entity-referenced sketch
constraints within parametric construction histories. We formulate a benchmark that compares
the construction and editing behavior of models built with alternative constraint sets.
We establish all-sketch constraint-generation baselines and examine how target-sketch
constraints, predictions in other sketches, training data, and auxiliary descriptions affect
model construction and edit success.

\section{Related Work}
\label{sec:related_work}

\paragraph{Constraint-based modeling and model reuse.}
Geometric constraint solving computes geometry satisfying specified relations
and dimensions~\citep{bettig2011geometric}. In parametric
modeling, the choice of constraints also expresses relationships to maintain
during revision~\citep{ault1999using,company2020constraints}. Model-reuse studies examine how reliably models can be modified under different modeling
strategies~\citep{camba2016reusability} and relate the use of the Fix constraint to the
effort required for later modifications~\citep{gonzalez2019fix}. Removing
redundant constraints has also been proposed to improve reuse~\citep{gonzalez2021redundancy}.

\paragraph{Learning sketch constraints.}
SketchGraphs represents sketches as geometric constraint graphs~\citep{sketchgraphs}, and HPSketch
includes histories with advanced sketch commands~\citep{hpsketch}.
Using SketchGraphs, Vitruvion generates primitives and constraints that can be
solved and edited in CAD software~\citep{vitruvion2022}. UniSketch combines sketch generation
and constraint prediction in a shared multi-task model~\citep{lin2026unisketch}.
Using solver feedback, \citet{Casey_2025_ICCV} encourage fully constrained,
solvable sketches without geometric distortion or over-constraint. AIDL embeds
constraint solving in a hierarchical language for LLM-driven generation and
editing of 2D designs~\citep{jones2025aidl}.

\paragraph{CAD histories and datasets.}
Evaluating sketch constraints in solid models also requires histories that retain the feature
operations using the solved geometry. The sequence representations in DeepCAD and Text2CAD encode sketches and
extrusions without explicit sketch constraints~\citep{wu_2021_iccv,khan2024text2cad}, whereas Fusion~360 Gallery
retains native sketch constraints and dimensions~\citep{willis2020fusion}.
WHUCAD includes additional types of feature operations and specifies which geometry each
feature selects~\citep{wos:001360811000005}. Program-based resources include executable
FeatureScript histories in CADFS~\citep{pyatov2026cadfs} and Python programs paired
with feature trees and validated geometry in FllumaOne~\citep{zhan2026fllumaone}.
HistCAD retains entity-referenced sketch constraints as learning targets together
with the feature operations needed to execute the construction history.

\paragraph{CAD editing and editability evaluation.}
CADMorph modifies construction sequences toward target geometry~\citep{cadmorph2025}, while
BenchCAD includes instruction-guided code editing~\citep{zhang2026benchcad} and
LLM4CAD-Editor studies editing at multiple program levels~\citep{sun2026llm4cadeditor}.
neuralCAD-Edit evaluates multimodal requests from professional designers using automatic
metrics and expert judgments~\citep{perrett2026neuralcadedit}. In these settings, the learned
system produces a modification in response to a request. In HistCAD, the predictor generates sketch constraints without receiving the edit request. The benchmark applies the
same prescribed dimensional edit to models built with alternative constraint sets.

MiCADangelo's constraint-editing experiment quantitatively compares reconstructed
and reference solids after the same sketch-point displacement, using individually
extruded SketchGraphs sketches~\citep{karadeniz2025micadangelo}. HistCAD applies dimensional
edits within retained construction histories and measures specified relations and unedited
dimensions, while requiring the initial model to be reproduced and the complete history
to rebuild successfully.

\section{HistCAD: Constraint-Aware Parametric Histories}
\label{sec:histcad-dataset}

HistCAD contains 180{,}495 parametric construction histories with 3{,}476{,}914 sketch
constraints in total. The histories use a common representation that combines initial
sketch geometry, entity-referenced geometric and dimensional constraints, and feature operations.

\subsection{Representation and Constraint Semantics}
\label{subsec:seq-format}

A HistCAD history is a nonempty ordered sequence of construction steps stored in JSON. A step
containing a sketch specifies its local coordinate system, primitives, constraints, and
associated feature operation. A step without a sketch specifies feature parameters and
references to existing geometry. Each sketch's constraint set is represented separately from
its primitive parameters and feature definition. It can therefore be replaced while these
definitions and the operation order remain unchanged.

Primitive parameters describe the initial geometry, whereas constraints specify the geometric
conditions the sketch solver must satisfy. For example, two lines may both be 20~mm long in
the initial sketch without being constrained to remain equal. An Equal constraint requires their lengths to remain equal during editing.
A Length constraint of 20~mm on one line, together with Equal, requires both lines to be 20~mm
long.

Sketches store primitives as a flat collection without explicit loop membership. Supported
primitives are lines, circles, circular arcs, ellipses, elliptical arcs, and non-uniform
rational B-spline (NURBS) curves. Constraints refer to these primitives or their points using
identifiers local to the sketch. Each primitive retains its identifier when its geometry
changes.

HistCAD defines 19 sketch-constraint types. The 12 geometric types specify relations between
entities or fix an entity or point. The seven dimensional types assign numerical values to
sizes, angles, and distances, including the major and minor radii of ellipses. Lengths are
expressed in millimeters and angles in degrees.

During construction, the sketch solver adjusts the initial geometry to satisfy the supplied
constraints. Feature regions are recovered from the solved primitives, and the associated
feature operation uses these regions to construct or modify a solid. Later operations can
refer to geometry created by earlier steps. Fillets and chamfers, for example, select edges
using stored query points whose coordinates remain unchanged during edited replay. An edit
can therefore leave an edge reference unresolved or ambiguous, preventing the affected feature
from executing.

\subsection{Dataset Construction and Coverage}
\label{subsec:histcad-sources}

Academic comprises 153{,}534 DeepCAD histories and 8{,}609 Fusion~360 Gallery histories,
while Industrial contains 18{,}352 histories (Table~\ref{tab:dataset-coverage}).
All histories are normalized to the same format before training and edit evaluation.

To add sketch constraints to DeepCAD histories~\citep{wu_2021_iccv}, we use source model or
sketch identifiers to pair them with SketchGraphs sketches~\citep{sketchgraphs} from the same
Onshape models. After DeepCAD post-processing,
geometric matching within each paired sketch identifies the HistCAD entities and points
corresponding to those used by the SketchGraphs constraints. We transfer the constraints by
rewriting their references with the mapped HistCAD identifiers. If any source constraint
supported by HistCAD contains an entity or point reference that cannot be mapped, the
entire history is excluded.

Fusion~360 Gallery histories~\citep{willis2020fusion} retain their original dimensions and
constraints. Industrial histories are extracted from native parametric CAD models from
multiple industrial sources. This conversion retains their geometric and dimensional sketch
constraints, feature operations, and geometric selections.

Industrial extends the extrusion-only coverage of DeepCAD and Fusion~360 to revolution,
helical sweep, fillet, and chamfer. Within this collection, 9{,}926 histories (54.1\%) combine
at least two operation families. Industrial histories are also longer on average than those
from DeepCAD and Fusion~360.

\begin{table}[!htbp]
\centering
\small
\caption{HistCAD collection sizes and history lengths. Constraint counts use the normalized entries and retain duplicate instances. Operation families are counted per collection. 95th pct. denotes the 95th percentile.}
\label{tab:dataset-coverage}
\setlength{\tabcolsep}{4pt}
\begin{tabular*}{\linewidth}{@{\extracolsep{\fill}}lrrrrrr@{}}
\toprule
\textbf{Collection} & \textbf{Histories} & \shortstack{\textbf{Sketch}\\\textbf{constraints}} & \shortstack{\textbf{Mean}\\\textbf{steps}} & \shortstack{\textbf{Median}\\\textbf{steps}} & \shortstack{\textbf{95th pct.}\\\textbf{steps}} & \shortstack{\textbf{Distinct op.}\\\textbf{families}} \\
\midrule
DeepCAD & 153{,}534 & 2{,}755{,}327 & 1.47 & 1 & 3 & 1 \\
\mbox{Fusion~360} & 8{,}609 & 155{,}603 & 2.23 & 2 & 6 & 1 \\
Industrial & 18{,}352 & 565{,}984 & 6.23 & 3 & 19 & 5 \\
\bottomrule
\end{tabular*}
\end{table}

For text-conditioned constraint prediction, the dataset also provides descriptions of model
construction, geometry, and possible use. They are generated from histories before sketch
constraints are removed to form predictor inputs, so they can mention those relations and
dimensions.

\section{Constraint-Preserving Editability Benchmark}
\label{sec:editability}
\label{sec:editability-benchmark}

For each history, the benchmark compares models built with different sketch-constraint sets
under the same dimensional edit. Initial primitive parameters, feature parameters, geometric
references, and operation order remain unchanged.

\subsection{Edit Tasks and Preservation}
\label{subsec:edit-tasks}

An edit task specifies a construction history, a target sketch, a dimensional change, and
the relations and unedited dimensions to preserve. The edited quantity is a line length,
distance, radius, or diameter, with initial and target values. A modification changes a
dimension specified by the history's constraints, whereas an addition assigns a dimension
not specified by them. Dimensions are considered equivalent when they specify the same
quantity on the same entities or points, with numerical conversion when needed.
Radius and Diameter, for example, describe a circle's size, with diameter twice its radius.
A modification requires exactly one equivalent dimension in the history, whereas an addition
requires none. Quantities with multiple equivalent dimensions in the history are excluded.
Other constraints may still restrict a quantity without a direct dimension.

Each variant is solved using its own constraints, but preservation is measured against the
same relations and unedited dimensions selected from the history's target sketch. For two lines required to remain
perpendicular, for example, the evaluator measures their angular deviation from $90^\circ$,
even if Perpendicular is absent from the prediction. A relation or dimension is selected
only if the evaluator implements its geometric measurement, its entity or point references
can be resolved, and its pre-edit deviation is finite. The dimension changed by a
modification is excluded.

\subsection{Constraint Variants}
\label{sec:constraint-completion}

Full Constraints (Full) retains all of the history's sketch constraints, which need not
fully constrain every sketch. These constraints form the reference set for task construction
and prediction scoring. To isolate the effect of the target sketch's constraints, Connectivity
retains only the Coincident constraints connecting profile endpoints in that sketch. Every
other sketch keeps Full's constraints.

All-sketch prediction uses predicted constraints in every sketch, evaluating their role
throughout the construction history. The predictor receives the history without its
explicit sketch constraints and is not given the edit request.

\subsection{Edit Execution and Outcome Metrics}
\label{subsec:executable-evaluation}
\label{subsec:driver-assisted}
\label{subsec:outcome-metrics}

\begin{figure}[!htbp]
\centering
\includegraphics[width=\widefigurewidth]{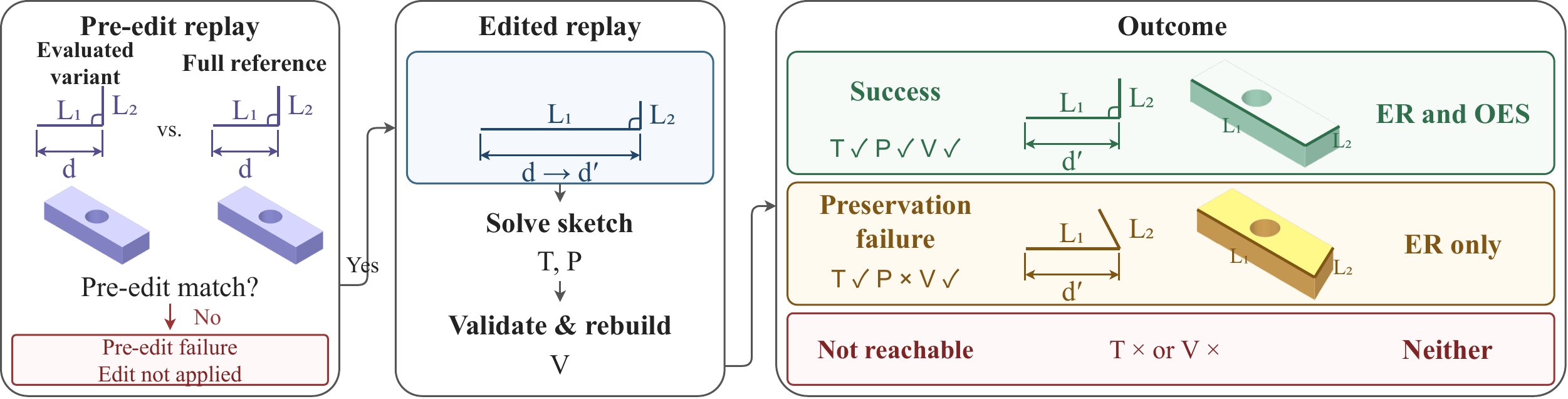}
\caption{Edit evaluation from initial-model reproduction to dimensional modification. ER is the fraction of assigned tasks whose models pass pre-edit matching, attain the target,
and rebuild validly (\(T=V=1\)). OES additionally requires preservation of every specified
relation and unedited dimension (\(P=1\)).}
\label{fig:editability-outcomes}
\Description{The workflow compares the evaluated pre-edit replay with the Full
reference, using both target-sketch and final-body geometry. Failure during
pre-edit replay or geometry matching prevents the edit. All target-sketch
insets identify the horizontal edited segment as L1 and the short edge
meeting its right endpoint as L2. Corresponding edges are highlighted and
labeled on the enlarged outcome renders. After matching, the request changes
the dimension from d to d prime; the target sketch is solved and T and P are
evaluated before continuing the target feature and downstream operations to
determine V. The green example is reachable and preserved, the yellow
reachable with preservation failure, and the red not reachable. The insets show one specified relation. P depends on all required relations and unedited
dimensions. The CAD bodies illustrate the outcome categories.}
\end{figure}

Before editing, each model must reproduce the geometry built under Full in
\mbox{Fusion~360}~\citep{fusion360API}. Before pre-edit replay for a modification, the evaluator
sets one equivalent dimension in the evaluated set to the history's pre-edit value. If none
is present, it inserts the designated dimension at that value. Additions leave dimensions
unchanged before replay, even if the prediction contains an equivalent dimension.

Pre-edit matching requires the prepared history to replay successfully and the resulting
solved target-sketch geometry to match that of the Full model within tolerance. The models must contain the
same number of final solid bodies, which are assigned one to one. For every assigned pair,
the bounding-box, relative-volume, and bidirectional sampled surface-distance residuals
must lie within their respective tolerances.

After pre-edit matching succeeds, the evaluator sets one equivalent dimension to the target
value or inserts the designated dimension at that value. All other constraints and feature
parameters remain unchanged, even when they conflict with the edit. Edited replay starts
again from the same stored initial primitives.

The evaluator measures the solved target sketch before its feature and later operations
execute. Target attainment ($T=1$) requires the measured edit quantity to match the requested
value within tolerance. For preservation ($P=1$), the deviation for each selected relation or unedited dimension
must be finite and within tolerance. Unedited dimensions are compared
with the values specified in the history. Unresolved geometric references and nonfinite
measurements count as failures. These measurements are retained even if a later feature fails.

Validity ($V=1$) requires every
supplied sketch constraint to be accepted, the target sketch to be valid and not over-constrained,
and every feature operation to execute successfully. \mbox{Under-constrained} sketches are permitted.
An edit
succeeds only if pre-edit matching passes and $T=P=V=1$ (Figure~\ref{fig:editability-outcomes}).

\noindent\textbf{Outcome metrics.}
For each constraint variant, $n$ counts all assigned tasks and $m$ those whose models pass
pre-edit matching. Among these, $r$ edits attain the target with a valid rebuild and are
termed \emph{reachable}. Of these, $s$ also satisfy preservation and are successful. Pre-edit
match rate $M_{\mathrm{pre}}$, edit reachability (ER), preservation rate (Pres.), and the
primary metric, overall edit success (OES), are
\begin{equation}
M_{\mathrm{pre}}=\frac{m}{n},\qquad
\mathrm{ER}=\frac{r}{n},\qquad
\mathrm{Pres}=\frac{s}{r},\qquad
\mathrm{OES}=\frac{s}{n}.
\label{eq:outcome-metrics}
\end{equation}
Both ER and OES use all assigned tasks, including failures before editing. Pres. uses the $r$ reachable edits as its denominator. It is undefined when $r=0$. For $r>0$,
$\mathrm{OES}=\mathrm{ER}\times\mathrm{Pres}$ when the rates are expressed as proportions.

\subsection{Task Construction and Selection}
\label{subsec:controlled-selection}

Edit tasks are selected before prediction. Before editing, models built under Full and
Connectivity must reproduce the same initial geometry, and their target sketches must satisfy
all selected relations and unedited dimensions. At least one selected relation or dimension
must derive from a non-Coincident constraint. Target values are sampled above or below the
Full model's pre-edit measurement with equal probability. Neither model may already attain the target
before editing.

After applying the candidate edit under Full, the target sketch must accept its constraints,
remain valid, reach the target, and preserve all selected relations and unedited dimensions.
Screening stops before the target feature, so failures from that feature onward remain in
the evaluation.

\section{Experimental Setup}
\label{sec:experimental-setup}

\subsection{Evaluation Sets}
\label{subsec:evaluation-sets}

The Academic and Industrial edit evaluation sets each contain 500 held-out histories,
with one edit per history. Each comprises 250 additions and 250 modifications selected
under Section~\ref{subsec:controlled-selection}. Source-model and design-variant groups stay
within one data split. The histories and edit requests are fixed before inference and shared
by all prediction conditions. Counts are summed across edit types before calculating rates.

\subsection{Predictors and Comparison Settings}
\label{subsec:prediction-models}

We fine-tune separate Qwen3-8B~\citep{qwen3} predictors with LoRA~\citep{hu2022lora} for
history-only and text-conditioned inputs. For each input type, Academic-only training uses
114{,}226 Academic histories, and joint training adds 15{,}768 Industrial histories to that
subset. This gives four trained predictors. Text-conditioned training presents each history
once with each of the Modeling Process (construction sequence), Geometric Feature, and
Functional Type descriptions. History-only training presents each history three times,
without the additional description tokens.

To test description--history correspondence, each text-conditioned predictor is evaluated
with a Modeling Process description from the evaluated history (same-history) or another
history in the collection's prediction test set (shuffled).
Pairings and structured histories are fixed, and paired inputs share an output-token cap for
each history. Together with the separately trained history-only predictor, these give three
conditions per training regime and six overall.

We separately test the influence of predictions outside the target sketch. Using same-history
outputs from both text-conditioned predictors, we keep the target-sketch prediction and use
the history's constraints in every other sketch (non-target restoration). The edit request,
preservation requirements, feature operations, and dimension-preparation rules remain
unchanged.

\subsection{Scoring and Statistical Analysis}

A response is available only if complete, parseable, and aligned with every sketch in the
input history. Otherwise its F1 is zero, and its task remains in the evaluation set as a
failure. Constraint validity and CAD replay are assessed separately from response structure.

Constraint F1 measures agreement with the history's constraints as the harmonic mean of
precision and recall. Matching is restricted to corresponding sketches and requires
the same constraint type, entity or point references, and nonnumeric attributes. Numerical
values must agree within the scoring tolerances. History-level F1 is computed from matched
and unmatched counts summed over the sketches of one history. Scoring uses the original
prediction, before the evaluator adjusts the edit dimension.

For history-only and same-history predictions, we relate history-level F1 to binary edit
success using Spearman's $\rho$. Because unavailable outputs contribute both zero scores
and failures, we also calculate the correlation using available responses, including those
whose models fail replay or geometry matching. For same-history predictions with
F1~$\geq0.8$, we count pre-edit failures, reachability failures, preservation failures,
and successful edits.

We use bootstrap resampling~\citep{efron1993bootstrap} to compute 95\% confidence intervals (CIs)
for OES and pre-edit matching differences and for F1 correlations. Industrial source-model or design-variant groups are resampled together.
Paired comparisons use the same resampled histories in both conditions. With the trained predictors fixed, the CIs quantify uncertainty from task sampling.

\section{Results}
\label{sec:experiments}

\subsection{Edit Success and Preservation}
\label{subsec:predicted-constraint-behavior}

\noindent\textbf{Generated constraints.}
Joint training with same-history descriptions gives the highest OES among the prediction
settings, reaching 52.4\% on Academic and 29.0\% on Industrial
(Table~\ref{tab:prediction-edit-results}). In this Industrial setting, 51.8\% of assigned tasks
pass pre-edit matching. The fraction that also reaches the target and rebuilds validly is
34.2\%, and requiring preservation gives 29.0\% overall success. Every prediction condition
includes models that reach the target and rebuild validly but fail preservation.

\begin{table}[!htbp]
\centering
\small
\caption{All-sketch prediction outcomes (\%). Joint denotes Academic--Industrial training. Maxima are bold for $M_{\mathrm{pre}}$, ER, and OES. The reachable edits included in Pres. can differ between conditions.}
\label{tab:prediction-edit-results}
\setlength{\tabcolsep}{2pt}
\begin{tabular*}{\linewidth}
{@{\extracolsep{\fill}}llrrrr@{\hspace{5pt}}rrrr@{}}
\toprule
& & \multicolumn{4}{c}{\textbf{Academic (500 tasks)}}
  & \multicolumn{4}{c}{\textbf{Industrial (500 tasks)}} \\
\cmidrule(lr){3-6}\cmidrule(lr){7-10}
\textbf{Training} & \textbf{Condition}
& $M_{\mathrm{pre}}$ & \textbf{ER} & \textbf{Pres.} & \textbf{OES}
& $M_{\mathrm{pre}}$ & \textbf{ER} & \textbf{Pres.} & \textbf{OES} \\
\midrule
Academic only & History only
& 50.8 & 44.4 & 87.8 & 39.0
& 25.8 & 20.4 & 70.6 & 14.4 \\
Academic only & Same-history description
& 59.8 & 55.8 & 91.8 & 51.2
& 28.2 & 24.6 & 80.5 & 19.8 \\
Academic only & Shuffled description
& 26.6 & 24.8 & 79.8 & 19.8
& 15.6 & 13.4 & 71.6 & 9.6 \\
\midrule
Joint & History only
& 52.6 & 45.8 & 88.2 & 40.4
& 48.8 & 29.2 & 76.0 & 22.2 \\
Joint & Same-history description
& \textbf{61.8} & \textbf{57.6} & 91.0 & \textbf{52.4}
& \textbf{51.8} & \textbf{34.2} & 84.8 & \textbf{29.0} \\
Joint & Shuffled description
& 25.6 & 23.6 & 85.6 & 20.2
& 31.6 & 22.0 & 79.1 & 17.4 \\
\bottomrule
\end{tabular*}
\end{table}

\noindent\textbf{Target-sketch controls.}
We compare Full and Connectivity in a separate control that retains the history's
constraints in every other sketch. More models reach the target
and rebuild validly under Connectivity, but fewer also preserve all required relations and
unedited dimensions (Table~\ref{tab:full-connectivity}). Compared with
Full, OES falls from 99.6\% to 53.0\% on Academic and from 79.2\% to 36.6\% on Industrial.
Full has 100\% Pres. among reachable edits because selection requires geometric
preservation in its edited target sketch.

\begin{table}[!htbp]
\centering
\small
\caption{Controlled edit outcomes (\%). Both variants have $M_{\mathrm{pre}}=100\%$ by task selection. Full's 100\% Pres. follows from the target-sketch feasibility criterion. $\Delta\mathrm{OES}$ denotes Full minus Connectivity in percentage points, with 95\% paired-bootstrap CIs.}
\label{tab:full-connectivity}
\setlength{\tabcolsep}{2.5pt}
\begin{tabular*}{\linewidth}{@{\extracolsep{\fill}}lrrrrrrr@{}}
\toprule
& \multicolumn{3}{c}{\textbf{Full}}
& \multicolumn{3}{c}{\textbf{Connectivity}} & \\
\cmidrule(lr){2-4}\cmidrule(lr){5-7}
\textbf{Task set}
& \textbf{ER} & \textbf{Pres.} & \textbf{OES}
& \textbf{ER} & \textbf{Pres.} & \textbf{OES}
& \shortstack{$\boldsymbol{\Delta}$\textbf{OES}\\
              \textbf{[95\% CI]}} \\
\midrule
Academic
& 99.6 & 100.0 & 99.6
& 100.0 & 53.0 & 53.0
& 46.6 [42.4, 50.6] \\
Industrial
& 79.2 & 100.0 & 79.2
& 82.4 & 44.4 & 36.6
& 42.6 [35.8, 49.6] \\
\bottomrule
\end{tabular*}
\end{table}

\FloatBarrier
\subsection{Influence of Constraints Outside the Target Sketch}
\label{subsec:non-target-constraint-behavior}
With the target-sketch prediction fixed, restoring the history's constraints in other
sketches enables additional models to reproduce the required initial geometry. Pre-edit
matching and OES improve in both collections under both training regimes
(Table~\ref{tab:prediction-diagnostics}). With joint training and same-history descriptions,
Industrial pre-edit matching rises from 51.8\% to 68.6\% and OES from 29.0\% to 41.0\%.

\begin{table}[!htbp]
\centering
\small
\caption{Edit outcomes before and after restoring the history's constraints outside the target sketch (same-history descriptions; rates in \%). Arrows denote all-sketch prediction $\rightarrow$ non-target restoration. OES uses all assigned tasks. Restoration adds reachable edits, so Pres. uses different subsets before and after restoration.}
\label{tab:prediction-diagnostics}
\setlength{\tabcolsep}{2pt}
\begin{tabular*}{\linewidth}{@{\extracolsep{\fill}}llrrrr@{}}
\toprule
\textbf{Task set} & \textbf{Training}
& $M_{\mathrm{pre}}$ & \textbf{ER} & \textbf{Pres.} & \textbf{OES} \\
\midrule
Academic & Academic only
& 59.8 $\rightarrow$ 74.8
& 55.8 $\rightarrow$ 68.6
& 91.8 $\rightarrow$ 92.1
& 51.2 $\rightarrow$ 63.2 \\
Academic & Joint
& 61.8 $\rightarrow$ 75.8
& 57.6 $\rightarrow$ 69.2
& 91.0 $\rightarrow$ 91.6
& 52.4 $\rightarrow$ 63.4 \\
\midrule
Industrial & Academic only
& 28.2 $\rightarrow$ 39.8
& 24.6 $\rightarrow$ 35.0
& 80.5 $\rightarrow$ 79.4
& 19.8 $\rightarrow$ 27.8 \\
Industrial & Joint
& 51.8 $\rightarrow$ 68.6
& 34.2 $\rightarrow$ 48.4
& 84.8 $\rightarrow$ 84.7
& 29.0 $\rightarrow$ 41.0 \\
\bottomrule
\end{tabular*}
\end{table}

Every additional success after restoration comes from a task whose model initially failed
pre-edit matching. Under joint training, 55 Academic and 60 Industrial tasks move from this
category to success. In these comparisons, models that already passed initial matching retain
their outcome categories.

\subsection{Auxiliary Descriptions and Industrial Training Data}
\label{subsec:auxiliary-conditions}

\noindent\textbf{Description--history correspondence.}
With the predictor and structured history fixed, same-history descriptions yield higher
OES than shuffled descriptions (Table~\ref{tab:prediction-edit-results}). Under joint
training, Industrial OES is 29.0\% versus 17.4\%, a difference of 11.6 percentage points
with a 95\% paired-bootstrap CI of [7.26, 16.20]. The 95\% CIs for these OES differences exclude zero in both collections under both
training regimes. Pre-edit matching and ER are
also higher with same-history descriptions in every comparison. The auxiliary information in
this comparison can include relations and dimensions, because the descriptions are generated
before constraint removal.

\noindent\textbf{Industrial training data.}
Joint training yields higher Industrial OES in every input condition
(Table~\ref{tab:prediction-edit-results}). With same-history descriptions, OES rises
from 19.8\% to 29.0\%, with a 95\% paired-bootstrap CI of [3.3, 15.9] percentage points
for the difference. This gain is accompanied by an increase in available predictions from
277 to 484 and in pre-edit matching from 28.2\% to 51.8\% over the same 500 histories. On Academic, the 95\% CIs
for OES differences between joint and Academic-only training include zero in every
input condition.

\subsection{Constraint F1 and Edit Outcomes}
\label{subsec:agreement-behavior}
History-level F1 is positively associated with edit success for history-only and
same-history predictions under both training regimes. Spearman's $\rho$ ranges from
0.473 to 0.605 on Academic and from 0.385 to 0.565 on Industrial, with every 95\% CI
excluding zero. This association persists among available predictions, including those whose
models fail initial replay or geometry matching.

\begin{table}[!htbp]
\centering
\small
\caption{Edit outcomes for predictions with history-level F1 $\geq0.8$, using same-history
descriptions. $|H|$ is the subgroup size. Membership can differ between predictors.
Counts are mutually exclusive and sum to $|H|$. Reachability failure means passing initial matching but failing target attainment or valid
rebuilding. Preservation failure means $P=0$ for a reachable edit.}
\label{tab:high-f1-stages}
\label{tab:f1-task-success}
\setlength{\tabcolsep}{4pt}
\begin{tabular*}{\linewidth}{@{\extracolsep{\fill}}llrrrrr@{}}
\toprule
Task set & Training & $|H|$ & \shortstack{Pre-edit\\failure} & \shortstack{Reachability\\failure} & \shortstack{Preservation\\failure} & Success \\
\midrule
Academic & Academic only & 282 & 52 & 10 & 15 & 205 \\
Academic & Joint         & 275 & 49 & 11 & 13 & 202 \\
Industrial & Academic only & 83 & 13 & 1 & 14 & 55 \\
Industrial & Joint         & 253 & 71 & 48 & 15 & 119 \\
\bottomrule
\end{tabular*}
\end{table}

Among tasks with high-F1 predictions, outcomes include successful edits and failures at
different stages of construction and editing (Table~\ref{tab:high-f1-stages}). With joint training and same-history descriptions,
predictions for 275 Academic and 253 Industrial tasks have F1~$\geq0.8$. Within these
subsets, 73 Academic tasks (26.5\%) and 134 Industrial tasks (53.0\%) are unsuccessful.
Of these failures, 49 on Academic and 71 on Industrial occur during initial replay or
geometry matching. Another 11 Academic and 48 Industrial tasks pass initial matching but
fail target attainment or valid rebuilding. A further 13 Academic and 15 Industrial models
reach the target and rebuild validly but fail the preservation criterion.

In one Academic example, a circle and an arc are required to retain a common center during
a diameter edit. The predicted constraints have a history-level F1 of 0.955. A model built
with these constraints reproduces the required initial geometry, reaches the requested
diameter, and rebuilds validly. After editing, however, the circle and arc centers are
0.289\,mm apart, exceeding the 0.01\,mm tolerance. The other 21 required relations and dimensions remain
within tolerance, but this separation makes the edit unsuccessful. The same edit succeeds
under Full.

\section{Discussion}
Under Connectivity, some relations and dimensions remain satisfied after editing even though
their corresponding explicit constraints are absent. Preservation requires every selected
relation and unedited dimension to pass its geometric check. In both the target-sketch
controls and all-sketch prediction results, some models reach the target and rebuild validly
while failing preservation.

The target sketch must also be considered within the complete construction history.
With the target-sketch prediction fixed, restoring the history's constraints in other sketches
yields additional successful edits, all from models that initially failed pre-edit matching.
After modification under Full, a target sketch can attain the requested dimension and
preserve all selected relations and unedited dimensions while the complete history fails to
rebuild. Edit success therefore depends on both the constraints used to construct the
required starting model and the feature operations that must execute after the target sketch
is modified.

The present experiments predict constraints while retaining the primitive and feature
definitions. The reported outcomes concern one prescribed edit per history. Future work could use construction and edit feedback
to guide the joint generation of sketch geometry, constraints, and feature operations.
Evaluating successive revisions could also test whether generated histories remain useful as
design requirements evolve.

\section{Conclusion}
HistCAD connects learning entity-referenced sketch constraints with constructing and editing
parametric CAD models through retained histories. Its benchmark compares alternative constraint
sets under the same prescribed dimensional edit while retaining the primitive and feature
definitions. Predictions outside the target sketch affect whether the required initial model
is reproduced. Target attainment and valid rebuilding do not guarantee preservation. Successful revision
depends on both the constraints governing geometry and the retained feature operations that
must execute after the target sketch is modified.
\FloatBarrier

\bibliographystyle{iclr2027_conference}
\bibliography{references}

\section*{Data and Code Availability}
\begingroup
\raggedright
The data are available at
\url{https://anonymous-hf.up.railway.app/a/i1n9x5t85qu5}.
The code is available at
\url{https://anonymous.4open.science/r/HistCAD-68C2}.
\par
\endgroup

\subsection*{AI Use Statement}
Generative AI tools were used to generate auxiliary textual descriptions of CAD histories
and to assist with code implementation following methods and implementation logic specified
by the authors. They also assisted with interpretation of the experimental results and with
organizing, drafting, and revising the manuscript. The authors take responsibility for the
final manuscript, data, and code.

\FloatBarrier
\appendix
\raggedbottom
\section{Representation and Data Construction}
\label{app:histcad-details}
\label{app:dataset-coverage}

\subsection{History Structure and Constraint Signatures}
\label{app:representation-schema}
\label{app:constraint-signatures}

All released histories use the same nonempty JSON-array representation, and the prediction
and edit evaluation experiments use this format after source normalization.
Each array element specifies one construction step (Table~\ref{tab:representation-schema}).
An entity identifier refers to a primitive in that step's sketch. Constraint references are
local to that sketch.

\begin{table}[!htbp]
\centering
\small
\caption{HistCAD history components used for constraint-aware replay and feature execution.}
\label{tab:representation-schema}
\setlength{\tabcolsep}{3pt}
\begin{tabularx}{\linewidth}{@{}p{0.22\linewidth}X@{}}
\toprule
\textbf{Component} & \textbf{Stored content and execution role} \\
\midrule
Coordinate system & Euler angles and a translation vector specify the pose of the local sketch plane in model coordinates. \\
Sketch primitives & Entity identifiers and line, circle, arc, ellipse, elliptical-arc, or NURBS parameters specify the stored sketch geometry. \\
Sketch constraints & Entries are grouped by type. For example, \texttt{Length: [[line\_1, 20.0]]} assigns a length to a line, while \texttt{Perpendicular: [[line\_1, line\_2]]} makes the two lines perpendicular. \\
Feature operation & Feature fields specify the operation family and its parameters, including extrusion extents, a revolution axis and angles, and helical pitch and turns. NewBody, Join, Cut, and Intersect specify body interaction. Fillets and chamfers specify edge references and size parameters. \\
Step order & The sequence determines which body geometry and references are available when each operation executes. \\
\bottomrule
\end{tabularx}
\end{table}

\paragraph{Local coordinates and primitive geometry.}
A sketch lies in its declared local XY plane. The \texttt{coordinate\_system} object
stores \texttt{Euler Angles} $(\alpha,\beta,\gamma)$ in degrees and a
\texttt{Translation Vector} $\mathbf t$ in millimeters. Using column vectors, the
coordinate transform is
\[
\mathbf p_{\mathrm{model}}=
R_x(\alpha)R_y(\beta)R_z(\gamma)
\begin{bmatrix}x\\y\\0\end{bmatrix}+\mathbf t.
\]
Horizontal, Vertical, ellipse-axis orientation, and directed Angle constraints are
interpreted in this declared frame. Geometric coordinates and feature parameters are
finite numerical values; dimension values may also use strings with explicit units,
such as \texttt{"20 mm"}.

The \texttt{sketch} object is a flat mapping from unique, type-prefixed identifiers to
primitive parameters. Table~\ref{tab:primitive-fields} specifies these fields. A circular
arc passes through \texttt{start}, \texttt{middle}, and \texttt{end}; its middle point
is neither its center nor necessarily its arc-length midpoint. The center is derived
from the nondegenerate three-point circle. Ellipse \texttt{major} and \texttt{minor}
values are semi-axis lengths, and \texttt{angle} specifies the major-axis direction
relative to the local X axis. For an elliptical arc, \texttt{large\_arc} selects an arc
larger than $180^\circ$ and \texttt{sweep=true} selects counterclockwise traversal.

\begin{table}[!htbp]
\centering
\small
\caption{Primitive fields in the common HistCAD format. A point is a two-element
coordinate array in the local sketch frame. Ellipse semi-axis lengths satisfy
\texttt{major} $\geq$ \texttt{minor} $>0$.}
\label{tab:primitive-fields}
\setlength{\tabcolsep}{4pt}
\begin{tabularx}{\linewidth}{@{}p{0.16\linewidth}p{0.21\linewidth}X@{}}
\toprule
\textbf{Primitive} & \textbf{Identifier prefix} & \textbf{Geometric fields}\\
\midrule
Line & \texttt{line\_} & \texttt{start}, \texttt{end}\\
Circle & \texttt{circle\_} & \texttt{center}, \texttt{radius}; optional \texttt{taper}\\
Circular arc & \texttt{arc\_} & \texttt{start}, \texttt{middle}, \texttt{end}\\
Ellipse & \texttt{ellipse\_} & \texttt{center}, \texttt{major}, \texttt{minor}, \texttt{angle}\\
Elliptical arc & \texttt{elliptical\_arc\_} & \texttt{start}, \texttt{end}, \texttt{major}, \texttt{minor}, \texttt{angle}, \texttt{large\_arc}, \texttt{sweep}\\
NURBS & \texttt{nurbs\_} & \texttt{degree}, \texttt{periodic}, \texttt{controls}; \texttt{weights} and \texttt{knots} subject to the conditions below\\
\bottomrule
\end{tabularx}
\end{table}

NURBS \texttt{degree} specifies the polynomial degree. Without \texttt{weights},
the curve is non-rational. When weights are supplied, they are finite and positive,
match the control-point count, and require an explicit knot vector. A supplied knot
vector contains $n+p+1$ nondecreasing finite values for $n$ control points and degree
$p$, with a nonempty active parameter domain. Periodic curves require an explicit
knot vector. Omitted knots use the execution adapter's default vector. Supplied weight and knot arrays
must be nonempty.

\paragraph{Canonical constraint syntax.}
The \texttt{constraints} object maps type names to arrays of entries with the
canonical forms in Table~\ref{tab:constraint-signatures}. An empty object denotes no
explicit constraints. An entity reference names a whole primitive, such as
\texttt{line\_1}; a point reference names one of its points, such as \texttt{line\_1.end}.

In the signatures, \texttt{p} denotes a point reference, \texttt{e} an entity reference,
\texttt{l} a line reference, and \texttt{c} a curve reference. The symbol \texttt{v}
denotes a finite value or an explicitly unit-bearing string. The Mirror operands
\texttt{o1} and \texttt{o2} may be points or entities, subject to the restrictions in the
table. The Distance arguments \texttt{ref1} and \texttt{ref2} identify the geometry used
by the distance form. The symbol \texttt{ellipse} denotes an ellipse entity. Major Radius
and Minor Radius use the keys \texttt{MajorRadius} and \texttt{MinorRadius}. The vocabulary
contains 12 geometric and seven dimensional types.

\begingroup\small
\begin{longtable}{@{}>{\raggedright\arraybackslash}p{0.18\linewidth}>{\raggedright\arraybackslash}p{0.38\linewidth}>{\raggedright\arraybackslash}p{0.38\linewidth}@{}}
\caption{Canonical constraint signatures. Each entry is stored in the array
under its type key. Referenced entities and points belong to the same sketch.}
\label{tab:constraint-signatures}\\
\toprule
JSON key & Single-entry form & Meaning and argument restrictions \\
\midrule\endfirsthead
\toprule JSON key & Single-entry form & Meaning and argument restrictions \\
\midrule\endhead
\bottomrule\endfoot
\texttt{Coincident} & \texttt{[p1, p2]} & Two points coincide. Both operands are point references. \\
\texttt{Concentric} & \texttt{[c1, c2]} & Two center-bearing curves share a center. \\
\texttt{Equal} & \texttt{[e1, e2]} & Compatible entities have equal corresponding sizes: line pairs, round-curve pairs, ellipse pairs, or elliptical-arc pairs. \\
\texttt{Fix} & \texttt{e} or \texttt{p} & An entity or point is fixed. \\
\texttt{Horizontal} & \texttt{l} or \texttt{[p1, p2]} & A line is horizontal or two points share a $y$ coordinate. \\
\texttt{Vertical} & \texttt{l} or \texttt{[p1, p2]} & A line is vertical or two points share an $x$ coordinate. \\
\texttt{Midpoint} & \texttt{[p, e]} or \texttt{[p, [p1, p2]]} & A point lies at the midpoint of a line, arc, elliptical arc, NURBS curve, or point pair. \\
\texttt{Mirror} & \texttt{[o1, [[x1,y1], [x2,y2]], o2]} & Two points or two entities of the same geometric type are reflected about the axis defined by the two distinct local points. \\
\texttt{Normal} & \texttt{[l, c]} (either order) & A line and a curve satisfy a normal relation. \\
\texttt{Parallel} & \texttt{[l1, l2]} & Two lines are parallel. \\
\texttt{Perpendicular} & \texttt{[l1, l2]} & Two lines are perpendicular. \\
\texttt{Tangent} & \texttt{[e1, e2]} & Two entities are tangent; at least one is a curve. \\
\texttt{Angle} & \texttt{[l1, l2, v]} & Angle between two ordered lines. \\
\texttt{Diameter} & \texttt{[c, v]} & Diameter of a circle or circular arc. \\
\texttt{Distance} & \texttt{[ref1, ref2, parameters]} & The parameter object contains \texttt{length} and may contain \texttt{direction}. \\
\texttt{Length} & \texttt{[l, v]} & Length of a line segment. \\
\texttt{Radius} & \texttt{[c, v]} & Radius of a circle or circular arc. \\
\texttt{MajorRadius} & \texttt{[ellipse, v]} & Major semi-axis of an ellipse. \\
\texttt{MinorRadius} & \texttt{[ellipse, v]} & Minor semi-axis of an ellipse. \\
\end{longtable}
\endgroup

The optional \texttt{Distance.direction} takes \texttt{MINIMUM},
\texttt{HORIZONTAL}, or \texttt{VERTICAL}; \texttt{length} is required.
Every Distance entry contains two references, and its parameter object contains only
these fields.

The two inline axis points in a Mirror entry are finite, distinct, and expressed in
local sketch coordinates. They define an infinite axis within the constraint entry.

Each Coincident, Parallel, Concentric, or Equal entry refers to exactly two points
or entities. Multi-member source relations are normalized before the histories enter
the dataset, training, or edit evaluation. The outer array under a type key collects
separate entries; for example, \texttt{Horizontal: ["line\_1", "line\_3"]} contains two separate single-line constraints. Scoring also defines how
to handle noncanonical entries produced by a predictor
(Section~\ref{app:content-evaluation}).

\subsection{Feature Regions and Geometric References}
\label{app:feature-regions}

\paragraph{Feature fields.}
Profile-driven features use regions recovered from the solved sketch, whereas fillets and
chamfers select edges of existing solids.
For profile-driven steps, feature parameters distinguish extrusion, revolution, and
helical sweep. Their \texttt{operation} value specifies \texttt{NewBody}, \texttt{Join},
\texttt{Cut}, or \texttt{Intersect}. Extrusion uses \texttt{towards} and \texttt{opposite}
for distances along and against the sketch normal. Revolution uses an axis point and
a nonzero direction, written as \texttt{[[px, py, pz], [dx, dy, dz]]}, and either
\texttt{start}/\texttt{end} or \texttt{start}/\texttt{angle} in degrees. Helical sweep
adds \texttt{pitch}, \texttt{turns}, and \texttt{handedness}, with handedness specified
as \texttt{Right} or \texttt{Left}. Fillet and Chamfer use their own operation values
and refer to existing edges using \texttt{near\_points}.

\paragraph{Boundary equivalence for selected planar faces.}
For a known selection of planar faces in the source model, HistCAD stores normalized
boundary primitives for their union. When distinct closed components intersect, normalization uses the source model to determine
an equivalent, unambiguous region interpretation.

\begin{figure}[!htbp]
\centering
\includegraphics[width=\linewidth]{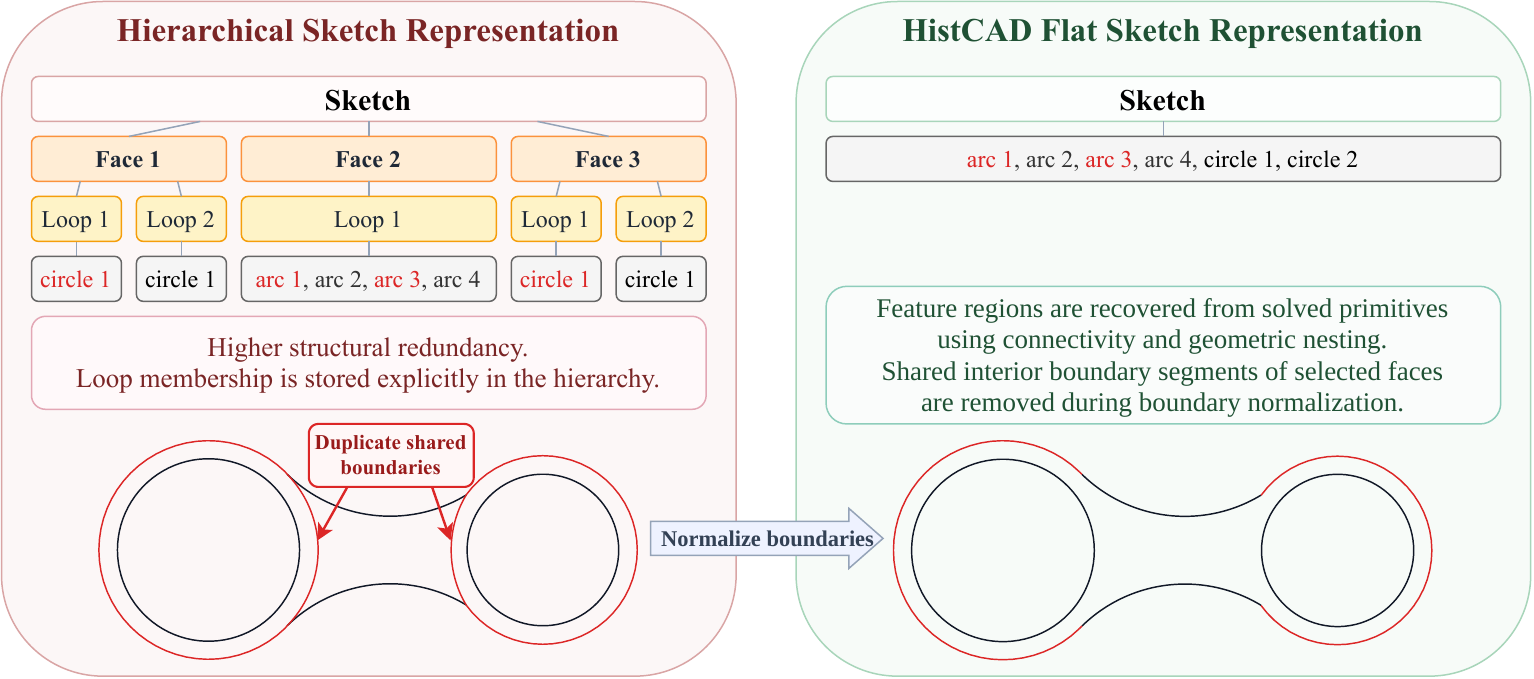}
\caption{Hierarchical and flat representations of the same selected sketch region. The
hierarchical form stores face--loop membership, whereas HistCAD stores normalized boundary
primitives. For the known face selection in the lower panels, shared interior boundary
segments cancel after splitting into atomic edges, leaving the exterior and hole boundaries.}
\Description{The upper panels compare a hierarchical sketch with explicit faces and loops
against a flat collection of boundary primitives. The lower panels show boundary
normalization for a known selection of planar faces. Shared internal boundary segments are
removed, while the exterior boundary and two circular holes remain. Feature regions are
recovered from solved primitives using connectivity and geometric nesting.}
\label{fig:flat-boundary-normalization}
\end{figure}

Let $F$ be a finite set of closed bounded faces in a planar arrangement, with pairwise
disjoint interiors. Split their boundaries into atomic edges at every intersection. Assume that an atomic edge is incident to at most two faces and is traversed at most once
by the boundary of any one face. Let $E(\partial f)$ denote the atomic-edge set of
$\partial f$, and let $\triangle$ denote symmetric difference. For
$U=\bigcup_{f\in F}f$, retain the edges with odd multiplicity:
\[
E_{\mathrm{flat}}=\mathop{\triangle}_{f\in F} E(\partial f).
\]
The geometric traces of these edges, together with their endpoints, form $\partial U$
(Figure~\ref{fig:flat-boundary-normalization}).

To see this, consider an interior point of one atomic edge and let $N$ be the number
of selected faces incident to it. If $N=0$, neither side belongs to $U$; if $N=2$,
both sides do. In neither case is the edge an exposed boundary of the selected region.
If $N=1$, one side belongs to $U$ and the other does not, so the edge is on its
boundary. Since $N$ is either 0, 1, or 2, odd multiplicity is equivalent to $N=1$.
Taking the endpoints completes the boundary. This construction removes shared internal segments from the stored boundary.

\paragraph{Region recovery during replay.}
During replay, the execution adapter recovers closed components from the solved versions
of the stored boundary primitives and determines the feature regions from their geometry. For nonintersecting closed
components, nesting depth alternates material and void: an outer component bounds material,
an immediately enclosed component bounds a hole, and the next enclosed component bounds a
material island. This nesting rule applies in either curve orientation.

When a single closed component intersects itself, curves are split at intersections,
and the bounded faces of the resulting planar arrangement are filled by the even--odd
rule within that component.

\paragraph{Point-based edge references.}

Fillet and chamfer operations identify edges using spatial query points and the geometry
available when that construction step executes. Edited replay retains the original query coordinates. A missing or ambiguous match to a
geometrically compatible edge leaves the reference unresolved and prevents the dependent
feature from executing.

\subsection{Source Pairing and Constraint Transfer}

Source model or sketch identifiers pair DeepCAD histories~\citep{wu_2021_iccv} with
SketchGraphs sketches~\citep{sketchgraphs} from the same Onshape model. Geometric matching
then relates each source sketch to the DeepCAD geometry after post-processing.
Establishing this correspondence can require splitting primitives at intersections or
merging collinear fragments. The resulting maps identify the HistCAD entities or points
that replace the source references in each transferred constraint. If any source constraint supported by HistCAD has a reference that cannot be mapped, the
entire history is excluded.

Fusion~360 Gallery histories~\citep{willis2020fusion} retain their source
constraints and dimensions. Industrial models originate from multiple industrial
sources, primarily from native parametric CAD files provided by companies.
Extraction retains primitives, constraints, dimensions, feature parameters,
geometric references, and operation order. Missing source constraints are not
inferred during conversion. Source-specific serialization is normalized to the common
format before the histories are used for training and edit evaluation.

\subsection{Dataset Statistics and Coverage}
\label{app:constraint-counting}

Explicit sketch constraints occur in 149{,}991 DeepCAD histories, 8{,}339 Fusion~360
histories, and 17{,}059 Industrial histories. These collections contain 17, 15, and 19
constraint types, respectively. Counting entries in the normalized histories, with
duplicate instances retained, gives 2{,}755{,}327 constraints for DeepCAD, 155{,}603
for Fusion~360, and 565{,}984 for Industrial.

\begin{figure}[!htbp]
\centering
\includegraphics[width=\widefigurewidth]{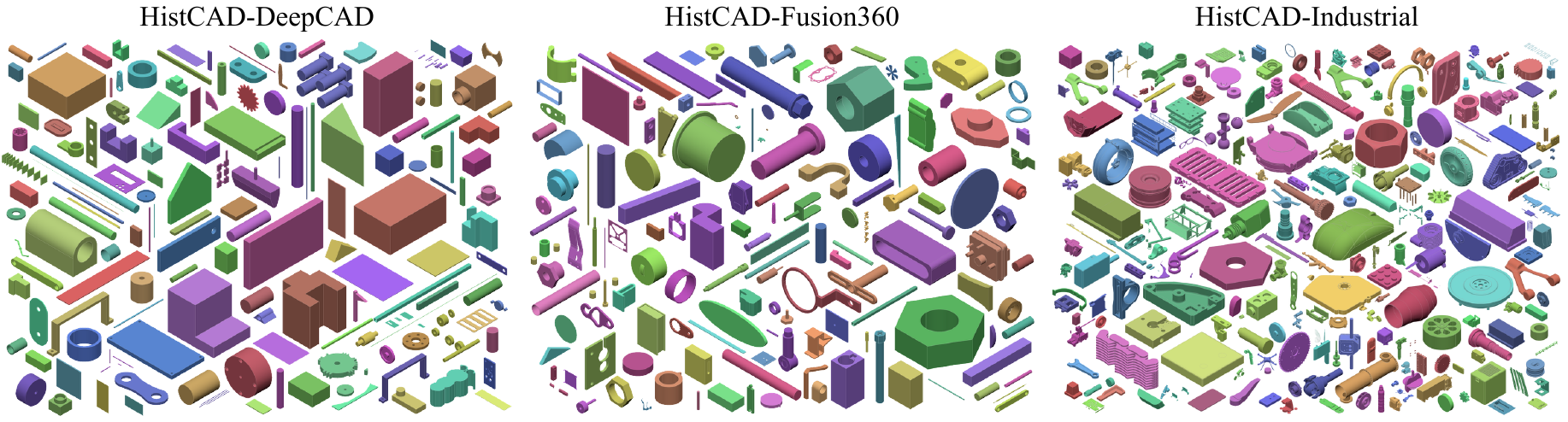}
\caption{Example final geometries from the DeepCAD, \mbox{Fusion~360}, and Industrial collections.}
\Description{Three galleries show final geometry examples from the DeepCAD, \mbox{Fusion~360}, and Industrial collections.}
\label{fig:dataset-complexity}
\end{figure}

Figure~\ref{fig:dataset-complexity} shows final geometry examples. Table~\ref{tab:industrial-feature-coverage}
reports history lengths, Industrial operation coverage, and the Industrial constraint
distribution. For operation coverage, a history containing several operation families
contributes to the history count for each family.

\begin{table}[H]
\centering
\small
\caption{Structural coverage and constraint counts. A: construction steps per history, using nearest-rank percentiles. A sketch and its feature may share a step. B: Industrial histories can contain multiple operation families. C: 565{,}984 Industrial constraint instances in normalized histories, with duplicate multiplicity retained.}
\label{tab:history-step-stats}
\label{tab:industrial-feature-coverage}
\setlength{\tabcolsep}{4pt}
\begin{tabular*}{0.88\linewidth}{@{\extracolsep{\fill}}lrrrr@{}}
\toprule
\multicolumn{5}{l}{\textit{Panel A: Construction-history length}} \\
\textbf{Collection} & \textbf{Histories} & \textbf{Mean} & \textbf{Median} & \textbf{95th pct.} \\
\midrule
DeepCAD & 153{,}534 & 1.47 & 1 & 3 \\
Fusion 360 & 8{,}609 & 2.23 & 2 & 6 \\
Industrial & 18{,}352 & 6.23 & 3 & 19 \\
\bottomrule
\end{tabular*}
\vspace{4pt}

\begin{tabular*}{0.72\linewidth}{@{\extracolsep{\fill}}lrr@{}}
\toprule
\multicolumn{3}{l}{\textit{Panel B: Industrial operation-family coverage}} \\
\textbf{Operation family} & \textbf{Steps} & \textbf{Histories} \\
\midrule
Extrude & 85{,}348 & 16{,}901 \\
Revolve & 9{,}181 & 5{,}372 \\
Helix sweep & 231 & 209 \\
Fillet & 13{,}341 & 5{,}040 \\
Chamfer & 6{,}284 & 3{,}958 \\
\midrule
At least two families & -- & 9{,}926 (54.09\%) \\
\bottomrule
\end{tabular*}
\vspace{4pt}

\begin{tabular*}{\linewidth}{@{\extracolsep{\fill}}lrrlrr@{}}
\toprule
\multicolumn{6}{l}{\textit{Panel C: Industrial constraint distribution}} \\
\textbf{Type} & \textbf{Constraints} & \textbf{\%} & \textbf{Type} & \textbf{Constraints} & \textbf{\%} \\
\midrule
Coincident & 327{,}662 & 57.89 & Concentric & 2{,}895 & 0.51 \\
Horizontal & 42{,}463 & 7.50 & Parallel & 2{,}704 & 0.48 \\
Perpendicular & 33{,}331 & 5.89 & Angle & 1{,}906 & 0.34 \\
Diameter & 30{,}962 & 5.47 & Fix & 1{,}830 & 0.32 \\
Tangent & 30{,}044 & 5.31 & Mirror & 883 & 0.16 \\
Vertical & 29{,}568 & 5.22 & Normal & 139 & 0.02 \\
Length & 29{,}090 & 5.14 & Midpoint & 91 & 0.02 \\
Distance & 13{,}226 & 2.34 & Minor Radius & 6 & $<$0.01 \\
Radius & 10{,}089 & 1.78 & Major Radius & 4 & $<$0.01 \\
Equal & 9{,}091 & 1.61 &  & &  \\
\bottomrule
\end{tabular*}

\end{table}

\subsection{Auxiliary Descriptions}
\label{app:text-annotations}

The three description types summarize the construction sequence (Modeling Process),
geometric features (Geometric Feature), and a possible use (Functional Type). Functional
Type is not an independently verified functional label. Gemma~4 31B IT~\citep{gemmateam2026gemma4}
generates these descriptions from histories before sketch constraints are removed to form
predictor inputs. The generation inputs contain primitives, geometric and dimensional
constraints, feature parameters, spatial information, and available model names.

All three descriptions are available for 153{,}518 DeepCAD, 8{,}609 Fusion~360,
and 18{,}320 Industrial histories, exceeding 99.8\% coverage in each collection.
Histories missing a description remain in HistCAD but are excluded from the
common text-comparison population.

\FloatBarrier
\Needspace{8\baselineskip}
\section{Dimensional Edits and Execution}
\label{app:experimental-details}

\subsection{Task Construction and Constraint Variants}
\label{app:behavior-protocol}
\label{app:task-construction}

Each selected history contributes one addition or modification task for a Length, Distance,
Radius, or Diameter dimension. For selecting an edit quantity, two dimensions are equivalent when they belong to the
same sketch-bearing step, describe the same quantity class, refer to the same entities or
points, and have matching direction or task-level half-space qualifiers where these apply.
A modification requires exactly one equivalent dimension in the history's original
constraint set. An addition requires none. Targets with multiple equivalent reference
dimensions are excluded.
Table~\ref{tab:canonical-quantities} specifies the canonical dimension forms used to identify
equivalent edit quantities.

\begin{table}[!htbp]
\centering
\small
\caption{Equivalent dimensions used to construct tasks and locate their targets.}
\label{tab:canonical-quantities}
\setlength{\tabcolsep}{3pt}
\begin{tabularx}{\linewidth}{@{}lXXl@{}}
\toprule
\textbf{Quantity class} & \textbf{Equivalent dimension forms} & \textbf{Target attachment} & \textbf{Addition form} \\
\midrule
Curve size & Radius or Diameter on the same circle or arc & curve identifier & Radius \\
Line size & Length on a line; or unsigned \texttt{MINIMUM} Distance between its own endpoints & line identifier & Length \\
Other distance & Directional, point--line, cross-entity, and other Distance constraints & canonical entity or point references plus direction and half-space attributes & canonical Distance form \\
\bottomrule
\end{tabularx}
\end{table}

Candidate histories containing line--circle or line--arc Normal constraints are excluded.
A candidate edit quantity must be measurable, refer to resolvable geometry, and have a
finite pre-edit value greater than $10^{-6}$\,mm. For a modification, the history's
dimension value must be within 0.01\,mm of the Full pre-edit measurement.

Candidates for additions include line lengths and the sizes of circles or circular arcs. Radius and Diameter candidates that identify the same curve size use a Radius target.
Distance additions are formed from line endpoints, circle centers, and arc endpoints and
centers. Pairs of points on different primitives define minimum, horizontal, or vertical
distances. Point--line pairs define minimum distance. Line--line Distance edits are limited to modifications of existing dimensions.

\paragraph{Candidate selection.}
The addition or modification category is assigned before the candidate edit is executed to
assess feasibility. The
candidate order alternates across sketches and quantity classes. For a Distance addition,
each point is paired with up to eight nearest points on other primitives and four nearest
other finite line segments. Point candidates are ranked by Euclidean distance, and line
candidates by point-to-segment distance.

One target value is sampled for each candidate. An increase or decrease is chosen with
equal probability, followed by a log-uniform scale factor from $[1/0.9,1/0.7]$ for an
increase or $[0.7,0.9]$ for a decrease. Multiplying the Full pre-edit measurement by this
factor gives the requested value.

\paragraph{Connectivity.}
Connectivity changes the constraint set only in the target sketch. It keeps Coincident
constraints between distinct endpoints of open curves. References to centers, interior
points, or endpoints of closed curves are excluded from this retained subset. A Coincident entry is retained only when both of its references are distinct eligible endpoints. Every other sketch
keeps its reference constraints. Connectivity uses the same edit-dimension preparation as
the other variants.

\paragraph{Selection criteria.}
Full and Connectivity must pass pre-edit matching, and their solved target sketches must
satisfy the specified relations and unedited dimensions before editing. For both Full and Connectivity, the difference between the pre-edit measurement and the
requested target must exceed twice the applicable outcome tolerance. The relations and dimensions to preserve must include at least one derived from a
non-Coincident reference constraint. Under Full, the edited target sketch must be valid,
accept its constraints, reach the target value, and preserve all these relations and dimensions. Screening ends before target-feature
execution.

We select 250 additions and 250 modifications separately for Academic and
Industrial. Academic candidate search alternates between DeepCAD and Fusion~360 histories
in the 6{,}382-history prediction-test set, while Industrial search uses 917
held-out histories (Section~\ref{app:evaluation-sets}).

\subsection{Dimension Preparation and Edited Replay}
\label{app:dimension-preparation}

The history's constraints determine whether the edit is an addition or modification.
Dimension preparation acts on the constraint set being evaluated. Equivalent dimensions
are identified using the quantity equivalence rules in Table~\ref{tab:canonical-quantities},
independently of predicted numerical values.

For a modification, the evaluator sets one equivalent dimension to the history's pre-edit
value $v_0$. If the evaluated set contains none, it inserts the designated dimension at
$v_0$. When several equivalents are present, one is selected deterministically and the
others remain unchanged. Selection prefers the type designated by the task. Within a
normalized type, candidates are ordered lexicographically by canonical content, then by
case-folded type keys and serialized original entries.

An addition receives no adjustment before pre-edit replay, even if the prediction contains
an equivalent dimension.

After pre-edit geometry matching succeeds, the evaluator updates one equivalent dimension
to the requested value or inserts the designated dimension. Radius--Diameter conversion
uses $d=2r$. Equivalent Length and Distance dimensions use the same numerical value.
All remaining constraints and later feature parameters are retained, including those that
conflict with the edit. F1 is calculated from the original prediction, before dimension
preparation or editing.

Pre-edit replay and edited replay are separate executions that both begin from the stored
primitives. Later sketches use the coordinate frames and primitives retained in their
history steps. Native implicit constraints and construction helpers are kept; helpers that
depend on the supplied constraints can differ between variants. Closed profiles are
recomputed from the solved sketch geometry.

For models built with predicted constraints, successful replay and geometry matching are
the pre-edit acceptance criteria. During edited replay,
the evaluator measures the edited quantity and the selected relations and unedited dimensions
after solving the target sketch, before its feature or later operations execute. These
measurements are retained even if a later feature fails.

Rebuilding is invalid if a supplied constraint is rejected or skipped, the target sketch is
invalid or over-constrained, a feature reference cannot be resolved, or an operation fails
or is skipped. Executable evaluation rejects predictions containing line--circle or
line--arc Normal constraints. This rejection causes execution failure without changing
structural availability or constraint F1.

\subsection{Geometry Matching and Preservation Evaluation}
\label{app:behavior-tolerances}

\paragraph{Pre-edit solved-sketch geometry.}
Local HistCAD references identify the native sketch entities to be measured. We compare
corresponding primitives after the sketch solver has computed their geometry, excluding
helpers generated by the CAD system. Matching requires corresponding
identifiers and primitive types. The geometric comparison uses ordered endpoints for lines,
centers and sizes for circles and ellipses, and the defining parameters of other primitive
types. Position and length differences are divided by $s_k=\max(10\,\mathrm{mm},d_k)$, where
$d_k$ is the bounding-box diagonal of the Full target sketch before editing. The tolerance
for these normalized differences is $10^{-6}$. Ellipse-axis orientation differences use a
tolerance of $10^{-6}$ radians, treating opposite directions along the same axis as
equivalent.

\paragraph{Final-body matching.}
The replayed model and the reference model must contain the same number of solid bodies. A
minimum-cost one-to-one assignment pairs bodies between the models using topological counts,
volume, and bounding-box geometry. Topological counts are used only to select the body pairs. Let
$s_b=\max(10\,\mathrm{mm},d_b)$, where $d_b$ is the diagonal of the bounding box enclosing
all final bodies in the Full model before editing. For each assigned
pair $B,B^0$, the bounding-box residual is the larger Euclidean distance between
the corresponding minimum and maximum corners, divided by $s_b$. The relative-volume residual is
$\lvert V(B)-V(B^0)\rvert/\max(\lvert V(B)\rvert,\lvert V(B^0)\rvert,10^{-15})$.
For the surface-distance residual, sampled points on $B$ are compared with the triangle
mesh of $B^0$,
and sampled points on $B^0$ are compared with the triangle mesh of $B$. In each direction,
the largest sampled point-to-mesh distance is retained. The surface residual is the larger
of these two distances divided by $s_b$. Every assigned body pair must satisfy all three
residual tolerances in Table~\ref{tab:preedit-settings}.
\begin{table}[!htbp]
\centering\small
\caption{Pre-edit geometry matching. The body residual tolerances apply to every pair in the
minimum-cost assignment. Topological counts are used only to select the body pairs.}
\label{tab:preedit-settings}
\begin{tabularx}{\linewidth}{@{}>{\raggedright\arraybackslash}p{0.25\linewidth}X>{\raggedright\arraybackslash}p{0.18\linewidth}@{}}
\toprule
Comparison & Quantity & Acceptance \\
\midrule
Solved-sketch position/length & Corresponding parameter residual divided by $s_k$ & $\leq10^{-6}$ \\
Solved-sketch orientation & Unoriented ellipse-axis angular residual & $\leq10^{-6}$ rad \\
Primitive identity & Corresponding identifiers and primitive kinds & Exact \\
Body count & Number of final bodies & Equal \\
Paired-body bounding box & Maximum corresponding-corner distance divided by $s_b$ & $\leq10^{-6}$ \\
Paired-body volume & Relative-volume residual & $\leq10^{-6}$ \\
Paired-body surfaces & Maximum bidirectional sampled point-to-mesh distance divided by $s_b$ & $\leq10^{-5}$ \\
\bottomrule
\end{tabularx}
\end{table}
\FloatBarrier

\paragraph{Preservation selection and measurement.}
Before editing, a relation or dimension is selected for preservation only if the evaluator
implements its geometric measurement, resolves the reference constraint's entity or point
references, and obtains a finite pre-edit residual (Table~\ref{tab:preservation-checks}).
A modification excludes the dimension being changed. The selection is identical for all
variants. Fix and Mirror remain in constraint scoring but are not measured for preservation.

After editing, the measured edit quantity is compared with the requested value to determine
target attainment. Each unedited dimension is compared with the value specified by its
reference constraint. An unresolved geometric reference or a nonfinite measurement gives a
failed result for the affected relation or dimension, which remains in the evaluation.
Table~\ref{tab:preservation-checks} gives the geometric quantities and tolerances.

\begin{table}[!htbp]
\centering
\footnotesize
\caption{Quantities and tolerances for target attainment and preservation.}
\label{tab:preservation-checks}
\setlength{\tabcolsep}{3pt}
\renewcommand{\arraystretch}{1.0}
\begin{tabularx}{\linewidth}{@{}l>{\raggedright\arraybackslash}Xl@{}}
\toprule
\textbf{Constraint type} & \textbf{Geometric quantity} & \textbf{Tolerance} \\
\midrule
Coincident & Maximum referenced-point separation & 0.01 mm \\
Concentric & Maximum center separation & 0.01 mm \\
Midpoint & Point-to-segment-midpoint separation & 0.01 mm \\
Parallel / Perpendicular & Line-direction deviation from \(0^\circ/180^\circ\) or \(90^\circ\) & \(0.5^\circ\) \\
Angle & Circular angular difference & \(0.5^\circ\) \\
Horizontal / Vertical & Line-to-axis angle; absolute \(y\)- or \(x\)-coordinate difference between the two referenced points & \(0.1^\circ\) for lines; \(\epsilon_\ell\) for points \\
Length / Radius / Diameter & Difference from the represented value & 0.01 mm \\
Major / Minor Radius & Difference from the represented ellipse semi-axis value & 0.01 mm \\
Distance & Point--point separation; point-to-supporting-line offset; offset between parallel supporting lines; absolute \(x\)- or \(y\)-difference for directional point pairs & 0.01 mm; \(\epsilon_\ell\) when directional \\
Equal & Difference in line length or curve radius & 0.01 mm \\
Tangent & Supporting-line--circle or circle--circle residual & 0.01 mm \\
\bottomrule
\end{tabularx}
\par\smallskip
\raggedright Neither evaluation set contains a Normal relation selected for preservation.
\end{table}

\Needspace{8\baselineskip}
\textit{Scale-aware tolerances.} For measurements assigned \(\epsilon_\ell\) in
Table~\ref{tab:preservation-checks}, the tolerance is \(\epsilon_\ell=10^{-4}s_\ell\), where
\(s_\ell\) is the maximum of 10 mm, the current evaluated sketch's bounding-box diagonal,
and the absolute length values used to evaluate the corresponding relation or dimension.
The scale is computed on the pre-edit target sketch during selection and on the solved
target sketch after editing during outcome evaluation.

\textit{Directional and line--line distances.} A line--line Distance is selected for
preservation only if the infinite lines supporting the two segments are parallel. After editing, the supporting lines must remain parallel within the angular tolerance
for their distance to count as preserved. Direction and task-level half-space qualifiers specify the sketch axis and geometric anchors used to compute the distance. These task qualifiers do not extend the canonical JSON parameter object.

\textit{Angle conventions.} Angle is measured counterclockwise from the first line's
start-to-end direction to the second line's direction in the declared local sketch XY
frame. Values wrap at \(360^\circ\); no line intersection is required. Degenerate line
directions are invalid.

\textit{Tangency conventions.} Tangency is evaluated on nondegenerate supporting curves, so a finite arc span need not contain the tangency point. Circular radii must be finite and positive. The line--circle residual is \(\lvert d(c,L)-r\rvert\), where \(c\) and \(r\) are the circle center and radius, and \(d(c,L)\) is the distance to the supporting line \(L\). For two circular curves with centers \(c_1,c_2\) and radii \(r_1,r_2\), it is \(\min(\lvert\|c_1-c_2\|-(r_1+r_2)\rvert,\lvert\|c_1-c_2\|-\lvert r_1-r_2\rvert\rvert)\). Coincident supporting circles of equal radius are treated as degenerate.

\paragraph{Per-constraint preservation outcomes.}
Each selected relation or dimension contributes one pass or failure to the per-constraint
totals. Duplicate reference constraints retain separate counts. An unresolved geometric
reference, a nonfinite measurement, or the absence of a target sketch yields a failure.
These measurement failures are distinguished from finite deviations outside tolerance.
Results from the solved target sketch are retained even when a later feature fails to rebuild.

\subsection{Editing with a Unique Predicted Dimension}

\label{app:unique-dimension-criterion}
The main protocol can insert a missing edit dimension or select among several equivalent
dimensions. The unique-dimension criterion additionally requires the original prediction
to contain exactly one equivalent edit dimension and the model to complete the edit under
the main protocol.

The evaluator still uses the history's pre-edit dimension value for initial replay and the
requested value for edited replay. Success rates use all 250 modification tasks as the
denominator. Unavailable predictions and predictions with zero or multiple equivalent
dimensions count as unsuccessful.

\FloatBarrier
\Needspace{8\baselineskip}
\section{Learning Setup and Statistical Methods}
\label{app:completion-details}

\subsection{Data Splits and Evaluation Populations}
\label{app:evaluation-sets}
\label{app:global-industrial-split}

In the common HistCAD format, construction steps are ordered, constraint values are finite,
and constraint references are local to their sketches. The prediction experiments use
histories with at least one sketch-bearing step, complete primitive and constraint entries,
and resolvable references. Within each sketch, canonical entries must have distinct combinations of constraint type,
referenced entities or points, and attributes. Dataset counts and generated predictions
retain duplicate entries. Successful Fusion~360 replay is not required for inclusion in the
prediction experiments.

\paragraph{Grouping and split assignment.}
In both the Academic and Industrial collections, histories associated with the
same source model or linked as design variants are assigned to the same split.

\paragraph{Academic population.}
Academic histories are divided into training, validation, and test sets in approximately
90/5/5 proportions. The split represents the source collections and histories with and
without explicit sketch constraints. Requiring every history to have all three
description types leaves 6{,}384 validation histories and 6{,}382 prediction-test histories.
The common training subset contains 114{,}226 histories whose history-only input and each
text-conditioned input fit within 8{,}192 tokens when the formatting and target response
are included. Inputs are used in full.

\paragraph{Industrial population.}
The Industrial partition contains 16{,}517 training histories, 918 validation histories, and
917 test histories. Requiring all three description types leaves 917 validation histories
and 915 prediction-test histories. Applying the common input-and-response token limit
retains 15{,}768 Industrial training histories, giving 129{,}994 histories in joint
training. Histories excluded from the prediction experiments remain in the released dataset.

\paragraph{Edit evaluation sets.}
Candidate search uses all 6{,}382 Academic prediction-test histories and the 917 held-out
Industrial histories. From these pools, 500 tasks per collection are selected before
inference. Every selected history has the descriptions and input lengths required by all
prediction conditions. The Academic set contains 280 DeepCAD and 220 Fusion~360 histories.
All 500 selected Industrial histories belong to the 915-history common prediction-test
population.

After the target sketch's step, 339 Academic histories have no subsequent construction
steps, 128 have one or two, 24 have three to five, and nine have at least six. The
corresponding Industrial counts are 151, 157, 98, and 94.

\subsection{Prediction Interface and Availability}
\label{app:prediction-interface}

Inputs are JSON histories with the supplied sketch geometry and feature operations retained
and all explicit sketch constraints removed. Text-conditioned inputs additionally contain a
description generated before constraint removal. Both same-history and shuffled evaluation
use Modeling Process descriptions; they differ in which history supplies the description.
The removed constraint sets provide training targets.

Training responses and inference outputs contain one constraint dictionary per sketch-bearing
step, in history order. Operations without sketches remain in the input but receive no
output dictionary. In this example for a history with one sketch-bearing step, Length
specifies the length of \texttt{line\_1}, and Perpendicular requires \texttt{line\_1}
and \texttt{line\_2} to be perpendicular. Both entries use identifiers from the input sketch.
\begin{lstlisting}
[
  {
    "Length": [["line_1", 20.0]],
    "Perpendicular": [["line_1", "line_2"]]
  }
]
\end{lstlisting}

A response is available only if it is complete, parseable, and aligned with all sketch-bearing
steps. Truncated or misaligned responses are rejected as a whole. An aligned
response can remain available even if individual entries have unknown types or malformed
contents. Parsing retains these entries, including repetitions, in their original order.
Constraint validity, CAD replay, and pre-edit geometry matching are assessed separately
from response structure.

\subsection{Training and Inference Configuration}
\label{app:auxiliary-text}

We train all four predictors with AdamW~\citep{loshchilov2019decoupled}, using the settings in Table~\ref{tab:training-configuration}. Text-conditioned training presents each eligible history once with each of the three
description types, without adding an explicit label for the description type. History-only
training presents the same histories three times. The predictors therefore see each history
equally often, although the text-conditioned inputs contain additional tokens. Only the
assistant-response tokens containing the history's constraints contribute to the training
loss.

\begin{table}[!htbp]
\centering
\small
\caption{Training and inference settings.}
\label{tab:training-configuration}
\setlength{\tabcolsep}{5pt}
\renewcommand{\arraystretch}{1.0}
\begin{tabularx}{0.92\linewidth}{@{}lX@{}}
\toprule
\textbf{Setting} & \textbf{Value} \\
\midrule
Base model & Qwen3-8B \\

Chat format and thinking mode & Qwen3 chat format; training and inference configured with thinking disabled \\
LoRA rank / alpha / dropout & 256 / 128 / 0.05 \\
LoRA modules & query/value projections \\
Optimizer & AdamW \\
Learning-rate schedule & Peak \(5\times10^{-5}\), 1\% warmup, cosine decay to 10\% of peak \\
Training epochs & History-only: 3; text-conditioned: 1 \\
Effective batch size & 16 \\
Context limit used & 8{,}192 tokens (prompt + output) \\
Requested output cap & 8{,}192 tokens, limited by remaining context \\
Minimum output window & 256 tokens \\
Training runs & 1 per predictor \\
Checkpoint rule & final training checkpoint \\
Training hardware & Four NVIDIA H800 GPUs per run \\
Decoding & greedy \\
\bottomrule
\end{tabularx}
\end{table}
\FloatBarrier

A shuffled description comes from another history in the same collection's held-out
prediction population and has the same description type as the corresponding same-history
input. Pairings are fixed. Donors may share a design family or similar wording with their
recipients.

For each history, we calculate the remaining context capacity separately for the same-history
and shuffled inputs. The two generations share an output-token cap equal to the smallest
of the requested limit and those two remaining capacities. Inputs are used in full. If the resulting cap is smaller than the minimum
output window, generation is skipped, but the input remains in the evaluation population.

\subsection{Constraint Matching and Scoring}
\label{app:content-evaluation}

Reference constraints use the common format. A generated response may contain
noncanonical entries: the scorer expands grouped Horizontal or Vertical line entries
and grouped Fix entries, retains grouped point entries, and counts duplicates separately. Entity and point references are canonicalized according to their roles in each
constraint. The order of symmetric references is ignored. Directional and other nonnumeric attributes in the constraint entries are retained
as supplied, and omitted default values are not inserted. Entity identifiers are not remapped by comparing geometry.

At corresponding sketch-bearing steps, a predicted and reference constraint can match only
if they have the same type, entity or point references, and nonnumeric attributes. Numerical
values use absolute tolerances of $10^{-3}$ mm and $10^{-3}$ degrees. Angular error is the
shortest circular difference under the directed-angle convention in
Section~\ref{app:behavior-tolerances}.
Swapping the ordered line references changes $\theta$ to $(-\theta)\bmod 360^\circ$.
For F1, Radius does not match Diameter, and Length does not match Distance, even when the
edit-construction rules treat them as equivalent dimensions. Malformed entries remain in
the scored prediction and cannot match valid reference constraints.

A maximum-cardinality one-to-one matching pairs as many predicted and reference entries
as possible, using each entry in at most one pair. Matched pairs count as true positives
(TP), unmatched predictions as false positives (FP), and unmatched reference entries as
false negatives (FN). F1 is $2\mathrm{TP}/(2\mathrm{TP}+\mathrm{FP}+\mathrm{FN})$ when
the denominator is nonzero. History-level F1 sums TP, FP, and FN across the sketches of one
history before applying this formula. Micro-F1 sums these counts across all histories before applying the formula. Non-Coincident
scores exclude Coincident constraints from the counts.

An unavailable prediction receives history-level F1 zero. Every reference constraint then counts as a false negative. If an available prediction and
its reference both contain no constraints, history-level F1 is one by convention. Such a
history adds no TP, FP, or FN to the pooled counts.

\subsection{Statistical Analysis}

We compute 95\% percentile CIs from 10{,}000 paired bootstrap resamples~\citep{efron1993bootstrap}. For Academic edit
tasks, individual histories are resampled within strata defined by source collection and
task type. For Industrial tasks, the resampling unit is a whole source-model or
design-variant group, with all its histories resampled together. Within each resample, we sum the task counts and recompute
$M_{\mathrm{pre}}=m/n$, $\mathrm{ER}=r/n$, $\mathrm{Pres}=s/r$, and $\mathrm{OES}=s/n$.   A paired difference
uses the same resampled task identities for both conditions. The tasks that pass matching or
become reachable can still differ between those conditions.

For correlation CIs, Academic histories are resampled within their source collections and
Industrial histories as whole source-model or design-variant groups. Ranks are recomputed in
each resample. The available-only analysis excludes unavailable outputs but retains predictions whose
models fail pre-edit matching. Correlations are undefined, for example, when all scores
or all outcomes in a resample are constant. The CI uses the defined correlations from the
original resamples. With the trained predictors fixed, these intervals quantify uncertainty
due to task sampling.

\FloatBarrier
\Needspace{8\baselineskip}
\section{Additional Results and Cases}
\label{app:additional-results}
\label{app:prediction-score-results}

\subsection{Constraint Scores on the Full Prediction Test Sets}
Constraint scores in Table~\ref{tab:completion-agreement-detailed} use the full prediction test sets of
6{,}382 Academic and 915 Industrial histories. Same-history descriptions yield higher pooled Micro-F1 than shuffled
descriptions under both training regimes in both collections. The same ordering holds when
Coincident entries are excluded. On Industrial, joint training also yields higher output
availability and Micro-F1 than Academic-only training. Unavailable outputs contribute
reference constraints as false negatives under the scoring rules in
Section~\ref{app:content-evaluation}.

\begin{table}[!ht]
\centering
\footnotesize
\caption{Constraint scores and availability on the full prediction test sets (\%). Scores pool TP, FP, and FN over each test set; Non-Coincident scores exclude Coincident constraints. P and R denote precision and recall. Available \(n\) (\%) counts complete, parseable, aligned outputs. Its denominator is all
6{,}382 Academic or 915 Industrial inputs, including those for which generation was skipped.
 }
\label{tab:completion-agreement-detailed}
\label{tab:constraint-agreement}
\label{tab:completion-output-quality}
\begin{tabular*}{\linewidth}{@{\extracolsep{\fill}}llrrrrr@{}}
\toprule
& & \textbf{All types} & \multicolumn{3}{c}{\textbf{Non-Coincident}} & \\
\cmidrule(lr){3-3}\cmidrule(lr){4-6}
\textbf{Training} & \shortstack[l]{\textbf{Prediction}\\\textbf{condition}} & \textbf{Micro-F1} & \textbf{P} & \textbf{R} & \textbf{F1} & \textbf{Available, \(n\) (\%)} \\
\midrule
\multicolumn{7}{l}{\textit{Academic prediction test set ($n=6{,}382$)}} \\
Academic only & History only & 67.40 & 40.88 & 53.73 & 46.43 & 6{,}300 (98.72) \\
Academic only & Same-history description & 72.98 & 51.57 & 60.88 & 55.84 & 6{,}299 (98.70) \\
Academic only & Shuffled description & 61.46 & 33.20 & 41.35 & 36.83 & 6{,}078 (95.24) \\
Joint & History only & 68.31 & 42.34 & 54.30 & 47.58 & 6{,}288 (98.53) \\
Joint & Same-history description & 73.01 & 51.88 & 60.44 & 55.84 & 6{,}299 (98.70) \\
Joint & Shuffled description & 62.74 & 34.52 & 41.80 & 37.81 & 6{,}011 (94.19) \\
\midrule
\multicolumn{7}{l}{\textit{Industrial prediction test set ($n=915$)}} \\
Academic only & History only & 35.58 & 32.40 & 18.78 & 23.78 & 476 (52.02) \\
Academic only & Same-history description & 37.49 & 41.52 & 20.33 & 27.30 & 481 (52.57) \\
Academic only & Shuffled description & 32.20 & 28.87 & 14.83 & 19.60 & 460 (50.27) \\
Joint & History only & 63.19 & 46.78 & 55.22 & 50.65 & 867 (94.75) \\
Joint & Same-history description & 64.55 & 49.22 & 55.54 & 52.19 & 871 (95.19) \\
Joint & Shuffled description & 58.94 & 41.75 & 47.33 & 44.37 & 829 (90.60) \\
\midrule
\end{tabular*}
\end{table}

\FloatBarrier
\Needspace{5\baselineskip}
\subsection{Paired Prediction Comparisons}
\label{app:joint-behavior-contrasts}

Table~\ref{tab:paired-prediction-contrasts} gives paired differences in both pre-edit
matching and OES for the same 500 tasks per collection. For joint minus Academic-only
training, the Industrial intervals are positive for both measures in every input condition,
whereas the Academic intervals include zero. For same-history minus shuffled descriptions,
all intervals are positive under both training regimes in both collections. Thus the
observed description-correspondence differences include initial-model reproduction as
well as complete edit success.

\begin{table}[!ht]
\centering
\footnotesize
\caption{Paired differences in pre-edit matching and overall success, expressed in percentage points with 95\% paired-bootstrap CIs. Each comparison uses the same 500 tasks per set. Panel headings specify subtraction order.}
\label{tab:paired-prediction-contrasts}
\setlength{\tabcolsep}{3pt}
\begin{tabular*}{\linewidth}{@{\extracolsep{\fill}}llrr@{}}
\toprule
\multicolumn{4}{l}{\textit{A. Joint minus Academic-only training}} \\
\textbf{Task set} & \textbf{Prediction condition} & \(\Delta M_{\mathrm{pre}}\) \textbf{[95\% CI]} & \(\Delta\mathrm{OES}\) \textbf{[95\% CI]} \\
\midrule
Academic & History only & $+1.80$ [-1.00, 4.80] & $+1.40$ [-1.40, 4.20] \\
Academic & Same-history description & $+2.00$ [-0.80, 5.00] & $+1.20$ [-1.60, 4.00] \\
Academic & Shuffled description & $-1.00$ [-4.00, 1.80] & $+0.40$ [-2.20, 3.00] \\
Industrial & History only & $+23.00$ [12.96, 32.64] & $+7.80$ [1.88, 14.42] \\
Industrial & Same-history description & $+23.60$ [15.07, 32.40] & $+9.20$ [3.30, 15.90] \\
Industrial & Shuffled description & $+16.00$ [9.72, 22.46] & $+7.80$ [3.91, 12.08] \\
\midrule
\multicolumn{4}{l}{\textit{B. Same-history minus shuffled descriptions}} \\
\textbf{Task set} & \textbf{Training} & \(\Delta M_{\mathrm{pre}}\) \textbf{[95\% CI]} & \(\Delta\mathrm{OES}\) \textbf{[95\% CI]} \\
\midrule
Academic & Academic only & $+33.20$ [28.20, 38.00] & $+31.40$ [26.80, 35.80] \\
Academic & Joint & $+36.20$ [31.40, 40.80] & $+32.20$ [27.60, 36.60] \\
Industrial & Academic only & $+12.60$ [8.51, 17.18] & $+10.20$ [7.21, 13.55] \\
Industrial & Joint & $+20.20$ [15.28, 25.28] & $+11.60$ [7.26, 16.20] \\
\bottomrule
\end{tabular*}
\end{table}

\FloatBarrier
\Needspace{5\baselineskip}
\subsection{Non-target Constraint Restoration}
Table~\ref{tab:completion-scope-flips} relates the increase in matched-model counts to the
additional successful edits after non-target restoration. The newly matched models have
both successful and unsuccessful edit outcomes. Every additional success comes from a
model that failed pre-edit matching under the original predictions. Models that already
passed matching retain their outcome category.

\begin{table}[!ht]
\centering
\footnotesize
\caption{Outcomes before and after restoring the history's constraints outside the target sketch,
using same-history descriptions and an unchanged target-sketch prediction. Both conditions
use the complete construction history. Arrows show before-to-after restoration values.
OES uses all 500 tasks in each set. The last column counts tasks changing from pre-edit
failure to success.}
\label{tab:completion-scope-flips}
\label{tab:local-completion-behavior}
\setlength{\tabcolsep}{3pt}
\begin{tabular*}{\linewidth}{@{\extracolsep{\fill}}llrrr@{}}
\toprule
\textbf{Task set} & \textbf{Training} & \textbf{Matched \(m\)} & \textbf{OES (\%)} & \shortstack{\textbf{Pre-edit failure}\\\textbf{$\rightarrow$ success}} \\
\midrule
Academic & Academic only & 299 $\rightarrow$ 374 & 51.2 $\rightarrow$ 63.2 & 60 \\
Academic & Joint & 309 $\rightarrow$ 379 & 52.4 $\rightarrow$ 63.4 & 55 \\
\midrule
Industrial & Academic only & 141 $\rightarrow$ 199 & 19.8 $\rightarrow$ 27.8 & 40 \\
Industrial & Joint & 259 $\rightarrow$ 343 & 29.0 $\rightarrow$ 41.0 & 60 \\
\bottomrule
\end{tabular*}
\end{table}

\FloatBarrier
\Needspace{5\baselineskip}
\subsection{Failure-Stage Accounting}
\label{app:failure-stage-accounting}

Each task is assigned to one of four mutually exclusive outcomes: pre-edit matching failure,
failure to reach the target or rebuild validly after matching, preservation failure among
reachable edits, or success. For $n$
assigned tasks, $m$ tasks that match, $r$ reachable edits, and $s$ successful edits, the
four counts are
\begin{equation}
n_{\mathrm{pre\text{-}fail}}=n-m,\quad
n_{\mathrm{reach\text{-}fail}}=m-r,\quad
n_{\mathrm{pres\text{-}fail}}=r-s,\quad
n_{\mathrm{success}}=s.
\label{eq:stage-counts}
\end{equation}
Here $n_{\mathrm{reach\text{-}fail}}$ counts tasks that pass matching but fail target
attainment or valid rebuilding.  Separate failure counts for $T$, $P$, and $V$ can overlap, and an early replay failure can
prevent later measurements or feature operations.

Applying Equation~\ref{eq:stage-counts} to the 500 tasks per condition gives the
outcome counts in Table~\ref{tab:all-task-stages}, corresponding to the rates in
Table~\ref{tab:prediction-edit-results}. Pre-edit failures are the largest failure category
in every prediction condition. The same partition applies within the high-F1 subsets in
Table~\ref{tab:high-f1-stages}.

\begin{table}[!ht]
\centering\small
\caption{Full-population outcome counts derived from Table~\ref{tab:prediction-edit-results}.
Every row sums to 500. Pre fail: unavailable prediction or failed pre-edit replay
or geometry matching; Reach fail: pre-edit matching passes but target attainment or valid rebuilding fails; Pres fail: reachable with $P=0$.}
\label{tab:all-task-stages}
\begin{tabular*}{\linewidth}{@{\extracolsep{\fill}}lllrrrr@{}}
\toprule
Task set & Training & Input & \shortstack{Pre\\fail} & \shortstack{Reach\\fail} & \shortstack{Pres\\fail} & Success \\
\midrule
Academic & Academic only & History only & 246 & 32 & 27 & 195 \\
Academic & Academic only & Same-history & 201 & 20 & 23 & 256 \\
Academic & Academic only & Shuffled & 367 & 9 & 25 & 99 \\
Academic & Joint & History only & 237 & 34 & 27 & 202 \\
Academic & Joint & Same-history & 191 & 21 & 26 & 262 \\
Academic & Joint & Shuffled & 372 & 10 & 17 & 101 \\
\midrule
Industrial & Academic only & History only & 371 & 27 & 30 & 72 \\
Industrial & Academic only & Same-history & 359 & 18 & 24 & 99 \\
Industrial & Academic only & Shuffled & 422 & 11 & 19 & 48 \\
Industrial & Joint & History only & 256 & 98 & 35 & 111 \\
Industrial & Joint & Same-history & 241 & 88 & 26 & 145 \\
Industrial & Joint & Shuffled & 342 & 48 & 23 & 87 \\
\bottomrule
\end{tabular*}
\end{table}

\FloatBarrier
\Needspace{5\baselineskip}
\subsection{Constraint F1 and Edit Success}
\label{app:correlation-results}
History-level F1 is positively correlated with edit success in every history-only and
same-history condition (Table~\ref{tab:agreement-behavior-association}A). This association
persists in the available-only populations in Panel B, which retain predictions whose
models fail initial replay or geometry matching. Panel A additionally counts failures
among predictions with F1 $\geq0.8$ and successes among those with F1 $\leq0.2$, separating
the overall association from outcomes at the two ends of the score range.

Each assigned task has binary outcome one for success and zero otherwise, including
failures before editing. High- and low-F1 subsets use scores over all assigned tasks;
unavailable predictions receive zero. Each reported fraction uses its own subset size
as denominator, and an empty subset has no defined rate.

\begin{table}[!ht]
\centering
\footnotesize
\caption{History-level F1 and edit success under all-sketch prediction. A uses all 500 tasks per set,
including pre-edit failures. For each predictor, failure and success fractions use the
numbers of tasks with prediction F1 $\geq0.8$ and $\leq0.2$, respectively, as denominators.
B excludes unavailable predictions but retains those whose models fail pre-edit matching.
Brackets give 95\% bootstrap CIs.}
\label{tab:agreement-behavior-association}
\setlength{\tabcolsep}{2pt}
\begin{tabular*}{\linewidth}{@{\extracolsep{\fill}}lllrrr@{}}
\toprule
\multicolumn{6}{l}{\textit{A. All  tasks}} \\
\textbf{Task set} & \textbf{Training} & \shortstack[l]{\textbf{Prediction}\\\textbf{condition}} & \textbf{$\rho$ [95\% CI]} & \shortstack[r]{\textbf{Failed tasks}\\\textbf{F1 $\geq0.8$}} & \shortstack[r]{\textbf{Successful tasks}\\\textbf{F1 $\leq0.2$}} \\
\midrule
Academic & Academic only & History only & 0.528 [0.455, 0.595] & 76/215 (35.3) & 0/12 (0.0) \\
Academic & Academic only & Same-history description & 0.595 [0.533, 0.653] & 77/282 (27.3) & 0/14 (0.0) \\
Academic & Joint & History only & 0.473 [0.394, 0.546] & 79/218 (36.2) & 0/12 (0.0) \\
Academic & Joint & Same-history description & 0.605 [0.543, 0.661] & 73/275 (26.5) & 0/13 (0.0) \\
Industrial & Academic only & History only & 0.416 [0.331, 0.501] & 17/47 (36.2) & 0/221 (0.0) \\
Industrial & Academic only & Same-history description & 0.565 [0.494, 0.633] & 28/83 (33.7) & 0/223 (0.0) \\
Industrial & Joint & History only & 0.385 [0.245, 0.510] & 141/222 (63.5) & 1/20 (5.0) \\
Industrial & Joint & Same-history description & 0.468 [0.346, 0.573] & 134/253 (53.0) & 0/16 (0.0) \\
\bottomrule
\end{tabular*}
\par\smallskip
\begin{tabular*}{\linewidth}{@{\extracolsep{\fill}}lllrr@{}}
\toprule
\multicolumn{5}{l}{\textit{B. Available predictions only}} \\
\textbf{Task set} & \textbf{Training} & \textbf{Prediction condition} & \textbf{Available \(n\)} & \textbf{$\rho$ [95\% CI]} \\
\midrule
Academic & Academic only & History only & 491 & 0.520 [0.445, 0.589] \\
Academic & Academic only & Same-history description & 487 & 0.579 [0.513, 0.641] \\
Academic & Joint & History only & 490 & 0.461 [0.380, 0.537] \\
Academic & Joint & Same-history description & 488 & 0.590 [0.526, 0.648] \\
Industrial & Academic only & History only & 279 & 0.216 [0.069, 0.381] \\
Industrial & Academic only & Same-history description & 277 & 0.423 [0.294, 0.538] \\
Industrial & Joint & History only & 481 & 0.374 [0.228, 0.503] \\
Industrial & Joint & Same-history description & 484 & 0.457 [0.331, 0.566] \\
\bottomrule
\end{tabular*}
\end{table}

\paragraph{High-F1 subgroup.}
For $H=\{i:F_{1,i}\geq0.8\}$, stage counts within $H$ separate pre-edit failures,
reachability failures, preservation failures, and successes into mutually exclusive
categories. Unavailable predictions have F1 zero and are absent from this subgroup.
Pre-edit failures here therefore occur with available predictions whose models fail initial
replay or geometry matching.

Among unsuccessful Academic tasks with same-history descriptions and high-F1 predictions,
pre-edit failures account for 67.5\% under Academic-only training and 67.1\% under joint
training (Table~\ref{tab:high-f1-stages}). Thus about two-thirds of the failures in both
Academic subsets occur before the requested edit. The corresponding Industrial proportions
are 46.4\% and 53.0\%. Each percentage describes the failure distribution within that predictor's own high-F1
subset. Subset membership can differ between predictors.

\FloatBarrier
\Needspace{5\baselineskip}
\subsection{Edit Types and Predicted Edit Dimensions}
\label{app:edit-type-dimension-results}

Under joint training with same-history descriptions, the 250 Academic additions and 250
Academic modifications each yield 131 successes, or OES of 52.4\%. Industrial additions
yield 68 successes (27.2\%), and modifications yield 77 (30.8\%), again out of 250 each.
The rates compare additions and modifications on different histories.

The unique-dimension criterion in Section~\ref{app:unique-dimension-criterion} requires the
original prediction to contain exactly one equivalent edit dimension and the model to
complete the edit under the main protocol. Predictions for 172 Academic modifications
contain exactly one equivalent dimension. Of the models built with these predictions,
132 pass pre-edit matching and 125 complete the edit. On Industrial, 158 predictions contain
exactly one equivalent dimension. Of their models, 101 pass matching and 66 complete the edit.

The unique-dimension criterion therefore gives 125 Academic and 66 Industrial successes,
compared with 131 and 77 under the main protocol. Both criteria use all 250 modifications
per collection. The six additional Academic successes and eleven additional Industrial
successes under the main protocol require insertion of a missing edit dimension.

\FloatBarrier
\Needspace{5\baselineskip}
\subsection{Additional Target-Sketch Constraint Comparisons}
\label{app:controlled-results}
An edit satisfies preservation only when all its selected relations and unedited dimensions
pass their geometric evaluations. A pooled rate counts individual preservation checks across histories. The Academic control set contains 4{,}918 individual checks and the Industrial set 4{,}813,
with duplicates counted separately. All pass under Full. Under Connectivity, 4{,}511
Academic checks (91.72\%) and 4{,}200 Industrial checks (87.26\%) pass. The counts retain
measurements from solved target sketches even if later rebuilding fails.

For Coincident, Connectivity passes 2{,}710 of 2{,}715 Academic checks and 2{,}583 of
2{,}597 Industrial checks. For other types, the corresponding counts are 1{,}801 of
2{,}203 (81.8\%) and 1{,}617 of 2{,}216 (73.0\%), although Connectivity omits those
constraints. These proportions use individual preservation checks as the denominator. On
Academic, the 407 failed checks are spread across 235 reachable edits, each with $P=0$.

Table~\ref{tab:paired-anchor-outcomes} follows the same tasks from Full to Connectivity.
On Industrial, five edits that are not reachable under Full succeed under Connectivity,
while eleven Full successes become not reachable. The table separates these transitions from preservation failures among
reachable edits.

\begin{table}[!ht]
\centering
\small
\caption{Paired Full--Connectivity outcomes on the complete controlled sets. Rows are
Full outcomes and columns are Connectivity outcomes. Each panel contains 500 paired tasks
from one collection. Row and column totals give the counts used for ER and OES. Both controls pass
pre-edit matching. Preservation failures include unresolved references and failed
measurements as well as measured geometric violations.}
\label{tab:paired-anchor-outcomes}
\begin{tabular*}{\linewidth}{@{\extracolsep{\fill}}lrrr@{}}
\toprule
\multicolumn{4}{l}{\textit{A. Academic}} \\
Full / Connectivity & Success & \shortstack{Reachable;\\preservation failure} & Not reachable \\
\midrule
Success & 264 & 234 & 0 \\
\shortstack[l]{Reachable;\\preservation failure} & 0 & 0 & 0 \\
Not reachable & 1 & 1 & 0 \\
\midrule
\multicolumn{4}{l}{\textit{B. Industrial}} \\
Full / Connectivity & Success & \shortstack{Reachable;\\preservation failure} & Not reachable \\
\midrule
Success & 178 & 207 & 11 \\
\shortstack[l]{Reachable;\\preservation failure} & 0 & 0 & 0 \\
Not reachable & 5 & 22 & 77 \\
\bottomrule
\end{tabular*}
\end{table}

\FloatBarrier
\Needspace{5\baselineskip}
\subsection{Constraint Prediction Comparisons}
\label{app:discordant-cases}
\begin{figure}[!htbp]
\centering
\includegraphics[width=\linewidth]{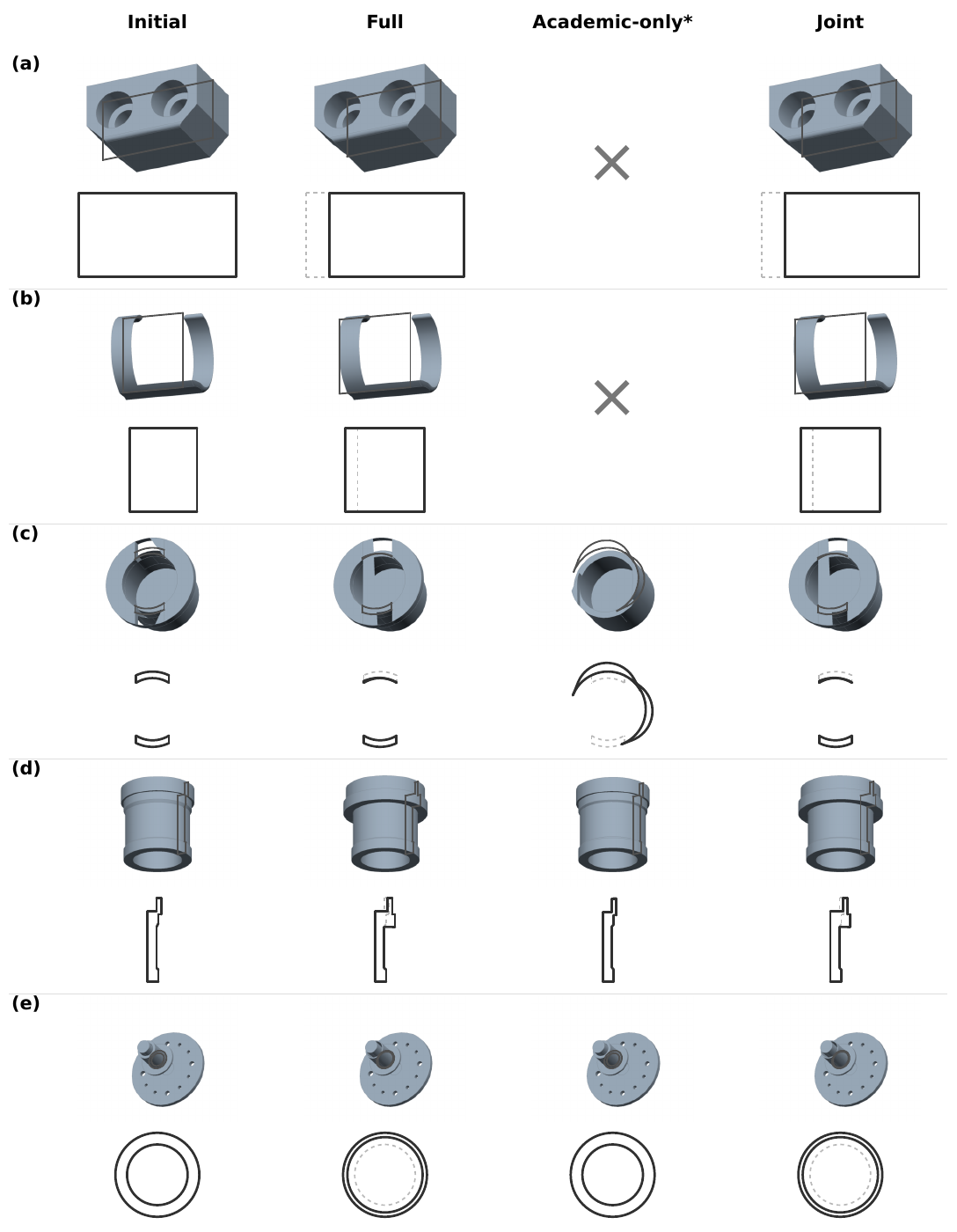}
\caption{Failures before the prescribed edit.
Initial is the reference. Full and Joint show successful edits.
Academic-only* fails before the requested change is applied.
The requests are (a) shortening the base of Industrial 01020012 (two chamfers
and a fillet) and (b) widening a cut profile in Industrial 01022672 (three fillets).
Extra predicted constraint groups prevent history construction in (a,b).
$\times$ marks these output errors.
In (c), the request is to increase the distance from an arc center to an endpoint
in Academic 01005050. During initial replay, Academic-only distorts the arc
profiles and fails Boolean extrusion (partial model shown).
The requests in (d,e) are to lengthen a sleeve-profile segment in Industrial
01027877 and enlarge the inner circle of Industrial 01016299, respectively.
The solved Academic-only sketches already mismatch the reference in both cases.
Models appear above target sketches. Dark solid lines show displayed geometry
and light dashed lines show the initial reference.}
\label{fig:prediction-before-edit}
\Description{Five comparisons with Academic-only failures before editing.
The first two rows mark Academic-only output errors with crosses; the
remaining rows show native pre-edit failures. Full and Joint complete the
edits in all five histories, including models with fillets and revolved features.}
\end{figure}

\begin{figure}[!htbp]
\centering
\includegraphics[width=\linewidth]{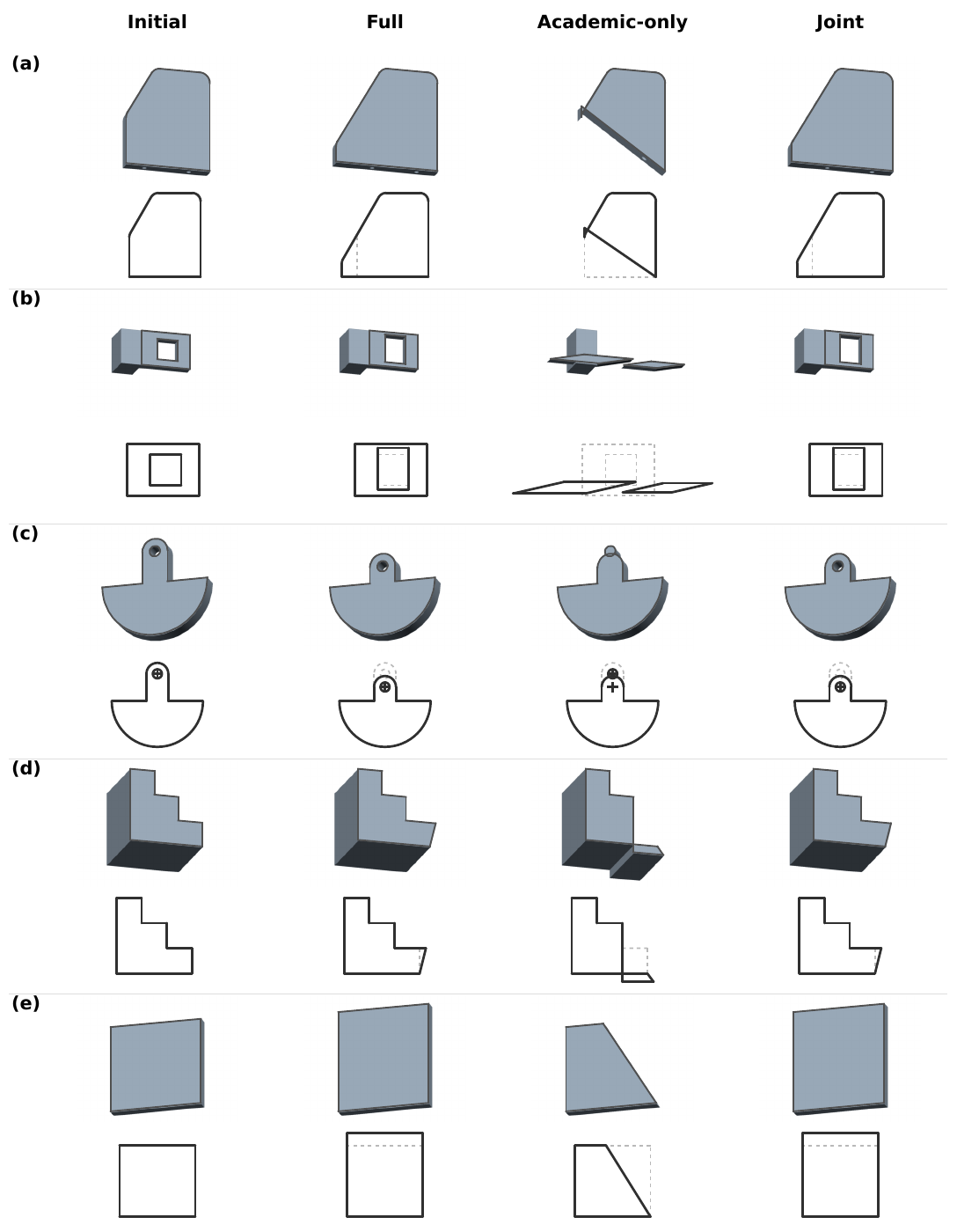}
\caption{Failures after pre-edit matching.
Initial is the reference; all other columns show results of the same edit.
Full and Joint succeed throughout.
Under Academic-only, (a) the lower edge of Industrial 01016487 reaches its
requested longer length, but a downstream cut misses the body;
(b) the opening in Academic 00465180 reaches its requested larger size, but a
perpendicular constraint is skipped and the sketch deforms. These two model
panels show incomplete replays.
In (c) Academic 01004348, reducing the distance across the lug leaves the hole
off-center within the rounded top.
In (d) Academic 00592341, extending a horizontal step edge also changes an
unedited edge length. In (e) Industrial 01024010, a specified right angle is
lost when one rectangle side is lengthened. The last three Academic-only results
attain the target and rebuild validly, but fail preservation.
Models appear above target sketches; dark solid lines show displayed geometry
and light dashed lines show the initial reference. Plus signs mark the centers
in (c).}
\label{fig:prediction-after-edit}
\Description{Five comparisons after applying the edit. The first two
Academic-only results fail valid execution. The other three attain the target
and rebuild validly but fail to preserve a concentric relation, an unedited
edge length, or a right angle. Full and Joint complete all edits successfully.}
\end{figure}

Figures~\ref{fig:prediction-before-edit} and~\ref{fig:prediction-after-edit}
compare four Academic and six Industrial tasks on which Joint and Full succeed
while Academic-only fails. Both predictors receive same-history descriptions
and predict constraints for every sketch. The two figures distinguish failures
before the prescribed edit from failures after pre-edit matching, when the edit
is applied.

Each row pairs native model renders with the corresponding target sketches.
All displayed sketch panels show native solved geometry. Thin gray outlines in
the model views locate the target sketch in its source plane.
Models share a camera and scale within each row; the sketch panels
likewise share coordinates and scale.

\end{document}